\documentclass[useAMS,usenatbib]{mn2e}

\usepackage{amsmath}
\usepackage{float}
\usepackage{morefloats}
\usepackage{graphicx}
\usepackage[english]{babel}
\usepackage{url}

\title[Collisional histories of early Vesta and Ceres]{Probing the history of Solar System through the cratering records on Vesta and Ceres}
\author[Turrini et al.]{D. Turrini$^{1}$\thanks{E-mail: diego.turrini@ifsi-roma.inaf.it}, G. Magni$^{2}$, A. Coradini$^{1}$\\
$^{1}$Institute for Interplanetary Space Physics, INAF, Via Fosso del Cavaliere 100, 00133, Rome, Italy\\
$^{2}$Institute for Space Astrophysics and Cosmic Physics, INAF, Via Fosso del Cavaliere 100, 00133, Rome, Italy}

\begin{document}

\date{Accepted XXX. Received XXX; in original form XXX}

\pagerange{\pageref{firstpage}--\pageref{lastpage}} \pubyear{2010}

\maketitle

\label{firstpage}

\begin{abstract}
Dawn space mission will provide the first, detailed data of two of the major bodies in the Main Asteroid Belt, Vesta and Ceres. Through its connection with HED meteorites, Vesta is known as one of the first bodies to have accreted and differentiated in the Solar Nebula, predating the formation of Jupiter and surviving the violent evolution of the early Solar System. The formation time of Ceres instead is unknown, but it should not postdate that of Jupiter by far, since the perturbations of the giant planet stopped planetary accretion in the Main Asteroid Belt. In this work we modelled the collisional histories of Vesta and Ceres at the time of the formation of Jupiter, assumed to be the first giant planet to form. In this first investigation of the evolution of the early Solar System, we did not include the presence of planetary embryos in the disk of planetesimals but we concentrated on the role of the forming Jupiter and the effects of its possible inward migration due to disk-planet interactions. Our results clearly indicate that the formation of the giant planet caused an intense early bombardment in the orbital region of the Main Asteroid Belt. We explored the effects of such bombardment on Vesta and Ceres assuming different size distributions of the primordial planetesimals. According to our results, Vesta and Ceres would not have survived the Jovian early bombardment if the disk was populated mainly by large planetesimals like those predicted to form in turbulent circumstellar disks. Disks dominated by small bodies, like those predicted to form in quiescent circumstellar disks, or with a varying fraction of the mass in the form of larger ($D\geq100$ km) planetesimals represent more favourable environments for the survival of the two asteroids. The abundance of planetesimals, especially the larger ones, proved a critical factor to this regards. The extent of Jupiter's radial migration due to disk-planet interactions proved itself another critical factor. In those scenarios where they survive, both asteroids had their surfaces saturated by craters as big as $150$ km and a few as big as $200-300$ km. In the case of Vesta, the Jovian early bombardment would have significantly eroded (locally or globally) the crust and possibly caused effusive phenomena similar to the lunar maria, whose crystallisation time would then be directly linked to the time of the formation of Jupiter.
\end{abstract}


\section{Introduction}\label{intro}

Our knowledge of the chronology and the evolution of the early Solar System is limited, particularly for what it concerns the first $10$ Ma. This is the timespan generally assumed as the upper bound to the lifetime of the gaseous component of the Solar Nebula and therefore to the formation of the giant planets. From the observations of circumstellar disks we know that their median lifetime is about $3$ Ma, with the range of observed values spanning between $1-10$ Ma \citep{hll01,mey08}. During this timespan, solid material should accrete to form the first planetesimals and then the planetary embryos from which the cores of the giant planets would originate. Such planetary cores should in fact appear in the forming Solar System early enough to allow for the accretion of the gaseous envelopes of the giant planets from the Solar Nebula.\\
The chronology of the early Solar System obtained through radiometric ages of chondrites, achondrites and differentiated meteorites indicates that the first solids to form, about $4567.2\pm0.6$ Ma ago, were the Ca-Al-rich inclusions \citep{aea02}, CAIs in the following. Chondrules, once thought to represent the oldest material that solidified in the Solar Nebula, formed about $1-3$ Ma later than CAIs (\citet{aea02}, \citet{cea08}) while differentiated bodies generally appeared in the next few million years after the formation of chondrules (see \citet{sco07} and references therein). However, meteoritic evidences suggest that in some cases differentiation of planetesimals took place extremely early in the history of the Solar System, i.e. about $1$ Ma after the formation of CAIs \citep{bak05,biz05}. Such primordial differentiation was due to the presence of short-lived radionuclides, mainly $^{26}$Al and $^{60}$Fe (see e.g. \citet{biz05}) in bodies larger than $20-30$ km in radius \citep{sco07}. In particular, the results by \citet{ygs07} obtained in studying the iron meteorites from the $IV A$ group suggest that these meteorites formed in a parent body that was about $300$ km wide and lacked an insulating mantle. The authors explained such anomalous composition of the parent body through the removal of the silicate-rich mantle from a differentiated protoplanet whose original size was about $10^{3}$ km (ibid). All these results collectively imply that planetary accretion started at the very beginning of the history of the Solar System and that a first generation of hundreds-of-km-wide bodies formed and differentiated in about $1-1.5$ Ma. Moreover, the study of differentiated meteorites (eucrites, ureilites and angrites, see \citet{sco07} and references therein) indicates that the differentiation of primordial planetesimals driven by short-lived radionuclides took place during a timespan covering the first $10$ Ma after the formation of CAIs.\\
The asteroid Vesta is of particular interest to this regard. Vesta has been identified as the possible parent body of the HED meteorites, a family of basaltic achondrites composed by howardite, eucrite and diogenite meteorites, and such connection would imply that the asteroid is differentiated (see e.g. \citet{dra01}, \citet{kei02} and references within). Moreover, the $^{40}$Ar-$^{39}$Ar ages of the oldest HED meteorites (see \citet{kei02}, \citet{sco07} and references within) suggest that this asteroid is primordial, i.e. it formed and differentiated in less than $4$ Ma since the formation of CAIs. In such scenario, Vesta would be the only known surviving primordially differentiated planetesimal and its formation would date back prior to or contemporary to the formation of the giant planets. As we previously mentioned, in fact, the giant planets should appear in the Solar System somewhere during the first $10$ Ma. The time needed for their formation is, to most practical purposes, the same over which they accrete their solid cores, estimated being of about a few Ma. The results of hydrodynamical studies in fact indicate that the phase of gas accretion took place at an extremely rapid pace. According to the simulations performed by \citet{lis09}, the total time it takes for Jupiter to accrete its gaseous envelope varies between several $10^4$ to a few $10^5$ years. In their simulations, \citet{cmt10} instead measured the time-scales of gas accretion for Jupiter and Saturn: their results vary between a few $10^3$ to about $10^5$ years depending on the physical parameters of the Solar Nebula. As a consequence of this match in the timing of the early differentiation of planetesimals and the formation of the giant planets, Vesta could bear the marks of Jupiter's birth.\\
The dwarf planet Ceres is of particular interest for a different reason. It has been suggested that, after the formation of the giant planets, the surviving planetary embryos influenced the evolution of the planetesimals in the Solar System and caused a depletion in mass \citep{wet92} in the orbital region of the Main Asteroid Belt respect to the mass hypothesised to originally reside in that region (see e.g. \citet{wei77}). The results by \citet{wet92}, \citet{caw01}, \citet{pmc01} and \citet{omb07} indicate that the combined gravitational perturbations of the giant planets and the planetary embryos reduce the population of planetesimals in which they are embedded by about a factor $100$ in about $10^{8}$ years. The planetary embryos themselves are removed on a $\sim 10^{7}$ years-long timescale by being ejected from the Solar System or being accreted by planetary bodies \citep{pmc01,omb07}. Being the most massive object which survived to present time in the Main Asteroid Belt, Ceres represent an important probe of the efficiencies and timescales of planetary accretion and removal in that orbital region.\\
In this first work we address the topic of the collisional evolution of primordial planetesimals during the formation of Jupiter, assumed to be the first giant planet to form in the Solar System. As we will describe, the formation of Jupiter causes a brief yet intense primordial bombardment in the Main Asteroid Belt. The bulk of this Jovian early bombardment lasts a few $10^{5}$ years, so our results describe a scenario where the formation of Saturn is delayed respect to that of Jupiter by about the same amount of time. In this first investigation of the evolution of the primordial Solar System, we did not take in account the possible effects of planetary embryos in the disk of planetesimals and, in particular, in the Main Asteroid Belt due to computational constrains. For the same reason, we did not consider the effects of the presence of the nebular gas on the orbital motion of the planetesimals. We will address the issues of the contributions of Saturn, the nebular gas and the planetary embryos in future papers.\\
Previous studies addressing the early evolution of the Solar System differ from our approach in several ways.
The studies dealing with the collisional history of the early inner Solar System adopted statistical approaches  designed to evaluate the evolution of the size-frequency distribution \citep{bea05a,bea05b,mea09} and the disruption law \citep{bea05a} of the planetesimals. We concentrated our attention on selected bodies, for which we reproduced possible early collisional histories. We chose Vesta and Ceres as our case studies since they will be visited in the next years by the Dawn space mission.
The studies investigating instead the dynamical evolution of early inner Solar System due to Jupiter's influence either used a sharp, step-like transition to describe Jupiter's formation \citep{pmc01} or directly assumed a fully formed Jupiter \citep{caw01,omb07}. In our study we concentrated on the temporal interval covering the accretion of the core and the gaseous envelope of the giant planet, trying to simulate this phase in a more realistic way, and took into account also the contribution of planetesimals from the outer Solar System.\\
The presentation of this work is organised in the following way. We describe in detail our physical and dynamical model in Section \ref{model}, while in Section \ref{results} we describe the output of our simulations. Finally, in Section \ref{implications} we discuss our results and their implications for the comprehension of the early Solar System and in Section \ref{conclusion}  we draw the conclusions about the interpretation of the collisional features that could be revealed by Dawn mission on Vesta and Ceres.

\section{Dynamical and physical model}\label{model}

To explore the early collisional history of Vesta and Ceres we simulated the dynamical evolution of a section of the young Solar System at the time of the formation of Jupiter's core and the subsequent accretion of its gaseous envelope. Our template of the forming Solar System was composed of the Sun, the accreting Jupiter, Vesta, Ceres and a swarm of massless particles representing the disk of planetesimals. We followed the evolution of our template of the Solar System for a temporal interval equal to $2\times10^{6}$ years. Vesta and Ceres were assumed already formed at the beginning of our simulations. The massless particles were initially distributed into a spatial region spanning between $2-10$ AU from the Sun. Such radial interval has been chosen after a set of numerical experiments to determine the region of the Solar System influenced by the forming Jupiter on the considered timespan, to optimise the usage of the computational resources. Being primarily interested to the effects of Jupiter's mass increase on the dynamical stability and the collisional evolution of the inner Solar System, we modeled the formation process of the giant planet through a semi-empirical approach. The timescales and the other parameters, on which our modeling was based, were derived from the results of hydrodynamical simulations described in \citet{cm04} and \citet{cmt10} and consistently with the findings of \citet{lis09}. In our semi-empirical model we considered also the effects of planetary migration due to the disk-planet interactions during the formation of Jupiter (see e.g. \citet{pea07} and references therein). During the dynamical evolution of our template of the Solar System we evaluated the probabilities of planetesimals impacting Vesta and Ceres through a statistical approach. In the following subsections we will describe in detail both the initial conditions and the physical parameters and constrains of our model.

\subsection{Vesta and Ceres}\label{asteroids}

Vesta and Ceres were assumed to be on circular orbits that lie on the same plane as that of Jupiter and do not change during the simulations. This approximation obviously neglects the orbital inclinations of the two asteroids, yet we adopted it since it significantly simplifies the treatment of their collisional evolution during the simulations, as described in Sect. \ref{impacts}. Moreover, we do not know if the present orbits of Vesta and Ceres are representative of their primordial ones.\\
The semimajor axes of Vesta and Ceres were obtained by the JPL Small-Body Database Browser\footnote{\url{http://ssd.jpl.nasa.gov/sbdb.cgi#top}}: the values adopted were respectively $a_{v}=2.362$ AU and $a_{c}=2.765$ AU. Their mean radii were assumed $r_{v}=258$ km \citep{tea97} and $r_{c}=476$ km \citep{tea05}. Finally, their mass values were derived from \citet{mic00} and were $m_{v}=2.70\times10^{23}$ g and $m_{c}=9.25\times10^{23}$ g. These values implies mean density values of $\rho_{v}=3.7\,g\,cm^{-3}$ \citep{mic00} and $\rho_{c}=2.0\,g\,cm^{-3}$ \citep{tea05}.

\subsection{Jupiter's formation and migration}\label{jupiter}

In our semi-empirical approach, we considered Jupiter's formation as composed by $3$ different stages:
\begin{itemize}
 \item a core accretion phase;
 \item a fast, exponential gas accretion phase;
 \item a slow asymptotic gas accretion phase.
\end{itemize}
The second and third stages of the formation process are ruled by the same physical process, i.e. Jupiter accreting gas from the Solar Nebula, and are analytically treated as a single stage in the model, yet they are clearly distinguishable from the point of view of the evolution of the early Solar System, as will be shown in Section \ref{results}.\\
At the beginning of the simulations, Jupiter is an embryo with mass $M_{0}=0.1\,M_\oplus$ and it grows to the critical mass $M_{c}=15\,M_\oplus$ in $\tau_{c}=10^{6}$ years. Since the total accretion time of Jupiter's core is the sum to our $\tau_{c}$ to the time needed to form the initial Mars-sized core, our choice of $\tau_{c}$ is consistent with the lower limits indicated by theoretical works for the formation of Jupiter's core (a few Ma, see e.g. \citet{nea07} and references therein). The mass growth is governed by the equation
\begin{equation}\label{coregrowth}
 M_{p}=M_{0}+\left( \frac{e}{e-1}\right)\left(M_{c}-M_{0}\right)\times\left( 1-e^{-t/\tau_{c}} \right)
\end{equation}
During most of the first stage, Jupiter's mass is negligible in terms of its perturbing effects on the planetesimals in the disk and the only planetesimals affected by Jupiter are those which undergo a close encounter with the forming planet. As a consequence, in this phase we expect that the number of impacts against the considered asteroids is governed by stochastic collisions of near-by objects.\\
As soon as Jupiter's core neared the critical mass value of $15\,M_\oplus$, the nebular gas surrounding it became gravitationally unstable and started to be rapidly accreted by the planet to form its massive envelope. During this phase, Jupiter directly perturbed near-by planetesimals, clearing a gap of increasing width in the disk, and more distant planetesimals through orbital resonances. During this phase, Jupiter's mass growth is governed by the equation
\begin{equation}\label{gasgrowth}
 M_{p}=M_{c}+\left( M_{J} - M_{c}\right)\times\left( 1-e^{-(t-\tau_{c})/\tau_{g}}\right)
\end{equation}
where $M_{J}=1.8986\times10^{30}\,g=317.83\,M_{\oplus}$ is Jupiter's final mass. As we previously anticipated, the timescale $\tau_{g}$ was derived from the hydrodynamical simulations described in \citet{cm04} and \citet{cmt10}. The value we employed in the simulation is $\tau_{g}=5\times10^3$ years, consistently with the findings of \citet{lis09}, and is linked to the turbulence parameter $\alpha=0.01$ used in modelling the disk.\\
In the third and final stage, Jupiter's accretion slows down while the giant planet reaches its final mass value and its gravitational perturbations secularly affect more and more distant planetesimals. As we discussed at the beginning of this subsection, during this phase Jupiter's mass growth is still governed by Eq. \ref{gasgrowth} and there is no discontinuity, from a numerical point of view, with the previous phase.\\
We followed the evolution of the template of the Solar System across the second and third stages for $\tau_{a}=10^{6}$ years, i.e. for $200\times\tau_{g}$. We wish to emphasise again that in our modelling Jupiter begins as a Mars-sized embryo. As a consequence, our starting time $t_{0}$ differs from the $t_{0}$ of the Solar System, i.e. the condensation of the first solids $4567.2\pm0.6$ Ma ago \citep{aea02}, and is located later, possibly by a few Ma. From a physical point of view, therefore, in our simulations we are looking at $2$ Ma-wide temporal windows located somewhere in the first $10$ Ma in the lifetime of the Solar System and centred on the time Jupiter started to efficiently capture the nebular gas.\\
In all our simulations, Jupiter starts on a circular orbit. However, theoretical models indicate that forming giant planets should undergo Type I and II migrations and drift inward due to their interactions with the protoplanetary disk (see e.g. \citet{pea07} and references therein). Type I migration is dominant while the mass of the forming planet is lower than $\approx100\,M_\oplus$ and the effects of the planet on the disk can be treated as linear perturbations (ibid). The timescale $\tau_{M}$ of Type I migration is a nonlinear function of the mass and the heliocentric distance of the forming planet: for a planet at $5.2$ AU, $\tau_{M}$ varies between $\sim10^{5}-10^{7}$ years \citep{dkh03}. Once the forming planet is massive enough to open a gap in the circumstellar disk, Type II migration takes over: the migration timescale becomes less sensitive to the planetary mass and for a planet at $5.2$ AU $\tau_{M}$ is of the order of $\sim10^{5}$ years (ibid). To evaluate the effects of the radial displacement of Jupiter on the early collisional histories of Vesta and Ceres, we forced an inward migration in Jupiter's motion. As a first approximation, we ignored the distinction between Type I and II migrations and started Jupiter's migration as soon as the mass of the forming planet reaches $15\,M_\oplus$, which is equivalent to say that the value of $\tau_{M}$ becomes of the order of $\sim10^{5}$ years (ibid). As a consequence, Jupiter moves on a circular orbit while its planetary core is growing to the critical mass and starts to spiral inward once the phase of gas accretion begins. The semi-empirical treatment of Jupiter's migration is analogous to Eq. \ref{gasgrowth} we used to describe the growth of the gaseous envelope:
\begin{equation}\label{radiallaw}
 r_{p}=r_{0}+\left( r_{J} - r_{0}\right)\times\left( 1-e^{-(t-\tau_{c})/\tau_{r}}\right)
\end{equation}
where $r_{0}$ is Jupiter's position at the beginning of the simulation, $r_{J}$ is the final position and $\tau_{r}=\tau_{g}=5\times10^{3}$ years. We assumed Jupiter's final semimajor axis equal to the present one, an assumption consistent with both the standard model of planetary formation and the scenario described by the so called Nice Model \citep{tsi05}. As a consequence, Jupiter's initial semimajor axis depends on the desired extent of radial displacement. In our simulations we considered four different migration scenarios: $0$ AU (no displacement), $0.25$ AU, $0.5$ AU and $1$ AU. The timescale $\tau_{r}$ we used in Eq. \ref{radiallaw} is an e-folding time while the timescale $\tau_{M}$ previously mentioned is defined as $\tau_{M}= a/\dot{a}$, so it represents the timescale for the planet to migrate from its initial position to the inner edge of the Solar Nebula. For an exponential decay law, about $95\%$ of the displacement is achieved in $3$ e-folding times, which in the case of Eq. \ref{radiallaw} is equal to $1.5\times10^{4}$ years. Therefore, displacements of $0.25$ AU, $0.5$ AU and $1$ AU with the assumed value of $\tau_{r}$ are equivalent to assuming values of $\tau_{M}$ at $5.2$ AU respectively of $\approx3.2\times10^{5}$ years, $\approx1.6\times10^{5}$ years and $\approx8\times10^{4}$ years, consistently with the results of theoretical studies \citep{pea07}.\\
Before proceeding with the description of the model, we would like to emphasise that the displacement we discuss here is not the one invoked by the Nice Model, which is temporally located several $10^{8}$ years later.

\subsection{Dynamical and physical characterisation of the planetesimals}\label{disk}

As anticipated, we modelled the dynamical evolution of the disk of planetesimals using a swarm of $8\times10^{4}$ massless particles. However, we associated to each particle a mass value and other physical features, which we used to model in a realistic way the effects of the impacts on Vesta and Ceres. The assumptions under which we derived the physical characteristics of the planetesimals will be detailed in the following.\\
The dynamical characteristics of the planetesimals in the disk at the beginning of our simulations are defined as follows:
\begin{align}
 2\,AU \leq &\,\,a_{i} \leq 10\,AU \nonumber\\
 0 \leq &\,\,e_{i} \leq 3\times10^{-2}\\
 0\,rad \leq &\,\,i_{i} \leq 3\times10^{-2}\,rad \nonumber
\end{align}
The values of eccentricity and inclination associated to each massless particle were chosen randomly as
\begin{equation}
 e_{i}=e_{0} X,\,i_{i}=i_{0} (1-X)%(1-2X)
\end{equation}
where $e_{0}=3\times10^{-2}$, $i_{0}=3\times10^{-2}$ rad and $X$ is a number extracted from a uniform distribution in the range $\left[0-1\right]$.\\
The planetesimals were first assumed to form by gravitational instability of the dust in the midplane of a nonturbulent protoplanetary nebula \citep{saf69,gaw73} having a mass of $M_{neb}=0.02$ M$_\odot$ distributed between $1-40$ AU with a density profile $\sigma=\sigma_{0}\left(\frac{r}{1\,AU}\right)^{-1.5}$ ($\sigma_{0}=2700$ g cm$^{-2}$ being the surface density at $1$ AU) and a dust to gas ratio $\xi=0.01$. Following \cite{cm81}, the mass spectrum of the planetesimals formed by the gravitational instability mechanism in this protoplanetary nebula spans the range $2\times10^{17}-10^{20}$ g. Different planetesimal formation mechanisms, however, would produce different size distributions and, as a consequence, different planetesimal abundances than the one we considered. As we will describe in Sect. \ref{turbulence}, therefore, we investigated the implications of planetesimal formation in turbulent disks by taking advantage of the results of \cite{mea09} and \cite{cha10}. While the results of \cite{mea09} directly supply the size-frequency distribution (SFD in the following) of planetesimals that formed in the region of the Main Asteroid Belt, we derived the size distribution and the abundance of the planetesimals in the formation scenario described by \cite{cha10} using the same analytical treatment we will now detail for the case of a quiescent disk.\\
From the results of \cite{cm81} we can derived the following semi-empirical relationship:
\begin{equation}\label{masslaw}
 \overline{m}_{p}=m_{0}\left( \frac{r}{1\,AU} \right)^{\beta}
\end{equation}
where $\overline{m}_{p}$ and $m_{0}$ are expressed in $g$, $r$ is expressed in $AU$ and $\beta=1.68$. The value $m_{0}$ is the average mass of a planetesimal at $1\,AU$, i.e. $2\times10^{17}$ g.\\
The number surface density in the disk of planetesimals can then be expressed as a function of mass and radial distance as
\begin{equation}\label{numdenslaw}
 n(m,r)=Q(r)m^{2}e^{-\left(m/\overline{m}_{p}(r)\right)^2}
\end{equation}
where $Q(r)$ represents the radial dependence of the number surface density $n(m,r)$ and we superimposed a Maxwell-Boltzmann distribution to the semi-empirical relationship of Eq. \ref{masslaw}.\\
The functional form of $Q(r)$ can be obtained by coupling Eq. \ref{numdenslaw} with the relationship governing the surface density profile in the Solar Nebula
\begin{equation}\label{nebdenslaw}
 \sigma_{p}(r)=\xi\sigma(r)=\xi\sigma_{0}\left( \frac{r}{1\,AU} \right)^{-n_s}
\end{equation}
where $\sigma_p$ is the surface mass density of the planetesimals, $\sigma$ is the surface mass density of the gas in the protoplanetary nebula, $\sigma_{0}=2700$ g cm$^{-2}$ is the surface mass density of the gas at $1\,AU$ and $\xi$ is the dust to gas ratio. As anticipated, we used the standard assumptions for the Solar Nebula, setting $n_s=1.5$ and $\xi=10^{-2}$. In the following, when the distance $r$ is implicitly normalised to $1$ AU it will be indicated with the capital letter $R$, while the symbol $1\,AU$ will indicate the value of the astronomical unit expressed in cm, i.e. $1\,AU=1.49597870691\times10^{13}\,cm$.\\
By integrating Eq. \ref{numdenslaw} over $m$ we obtain
\begin{align}\label{numlaw}
 n^{*}(r)=\int^{\infty}_{0}n(m,r)dm= \nonumber \\
=\int^{\infty}_{0}Q(r)m^{2}e^{-\left(m/\overline{m}_{p}(r)\right)^2}dm=\frac{\sqrt{\pi}}{4}Q(r)\overline{m}_p(r)
\end{align}
while for the surface mass density of planetesimals we have
\begin{align}\label{surfdenslaw}
 \sigma_{p}(r)=\int^{\infty}_{0}n(m,r)mdm= \nonumber \\
=\int^{\infty}_{0}Q(r)m^{3}e^{-\left(m/\overline{m}_{p}(r)\right)^2}dm=\frac{1}{2}Q(r)\overline{m}_{p}^{4}(r)
\end{align}
By equating Eq. \ref{surfdenslaw} with Eq. \ref{nebdenslaw} we get
\begin{equation}\label{qlaw}
 Q(r)=2\frac{\xi\sigma_{0}}{\overline{m}^{4}_{p}}R^{-n_{s}}
\end{equation}
which, substituting Eq. \ref{qlaw} into Eq. \ref{numlaw} and applying Eq. \ref{masslaw}, gives
\begin{equation}\label{nstar}
 n^{*}(r)=\frac{\sqrt{\pi}}{2}\frac{\xi\sigma_{0}}{m_{0}}R^{-(n_{s}+\beta)}
\end{equation}
Now we can obtain the cumulative distribution of $n^{*}(r)$ by integrating
\begin{align}\label{cumnum}
 N(x)=\int^{x}_{r_{min}} 2 \pi r n^{*}(r)dr= \nonumber \\
 =\pi^{3/2}\frac{\xi\sigma_{0}}{m_{0}}\left(1\,AU\right)^{2}\left(\frac{1}{2-n_{s}-\beta}\right)\times \nonumber \\
 \left(\left(\frac{x}{1\,AU}\right)^{2-n_{s}-\beta}-\left(\frac{r_{min}}{1\,AU}\right)^{2-n_{s}-\beta}\right)
\end{align}
where $r_{min}=2$ AU is the inner radius of the disk of planetesimals.
We can obtain the total number of planetesimals in the disk by integrating Eq. \ref{cumnum} between $r_{min}=2$ AU and $r_{max}=10$ AU:
\begin{align}\label{ntot}
 N_{tot}=\int^{r_{max}}_{r_{min}} 2 \pi r n^{*}(r)dr= \nonumber \\
 =\pi^{3/2}\frac{\xi\sigma_{0}}{m_{0}}\left(1\,AU\right)^{2}\left(\frac{1}{2-n_{s}-\beta}\right)\times \nonumber \\
 \left(\left(\frac{r_{max}}{1\,AU}\right)^{2-n_{s}-\beta}-\left(\frac{r_{min}}{1\,AU}\right)^{2-n_{s}-\beta}\right)
\end{align}
The function $X=\frac{N(x)}{N_{tot}}$ uniformly varies in the range $0 \leq X \leq 1$. Once we invert the previous equation and express the initial position of the planetesimals in the disk as $x=f(X)$ where $\mu=2-n_{s}-\beta$ and
\begin{align}
 x=f(X)= \nonumber \\
 \left[X\left(\left(\frac{r_{max}}{1\,AU}\right)^{\mu} -
 \left(\frac{r_{min}}{1\,AU}\right)^{\mu}\right)+\left(\frac{r_{max}}{1\,AU}\right)^{\mu}\right]^{-\mu}
\end{align}
we can use $X$ as the random variable in a Montecarlo extraction to spatially populate the disk.\\
Through Eq. \ref{masslaw} we can link the average mass $\overline{m}_{p}(r)$ of the planetesimals to their initial position $r=x$ (which we implicitly assume coinciding with the formation region). To obtain the mass value of each planetesimal, we apply again a Montecarlo method through a uniform random variable $Y$. To obtain $Y$ we need to compute and normalise the cumulative distribution of Eq. \ref{numdenslaw} considering $m$ as the variable of interest and $r$ as a constant. The cumulative distribution is
\begin{equation}\label{cummass}
 \int^{m^{*}}_{0}n(m,r)dm=\frac{2\xi\sigma_{0}}{\overline{m}_{p}(r)}R^{-n_{s}}\int^{y^{*}}_{0}y^{2}e^{-y^{2}}dy
\end{equation}
where $y=m/\overline{m}_{p}(r)$ and $y^{*}=m^{*}/\overline{m}_{p}(r)$.\\
The integral on the right side of Eq. \ref{cummass} cannot be solved analytically: substituting $z=y^{2}$ with $dz=2ydy$ we obtain
\begin{equation}
 \int^{y^{*}}_{0}y^{2}e^{-y^{2}}dy=\frac{1}{2}\int^{\sqrt{z^{*}}}_{0}z^{1/2}e^{-z}dz
\end{equation}
The integral on the right side of previous equation is a lower incomplete Gamma function with real parameter $a=3/2$ or $\gamma\left(3/2,\sqrt{z^{*}}\right)$.
Normalising Eq. \ref{cummass} over Eq. \ref{nstar} we get the uniform random variable $Y$ varying in the range $[0,1]$
\begin{equation}
 Y=\frac{2\gamma\left(3/2,y^{*}\right)}{\sqrt{\pi}}=P\left(3/2,y^{*}\right)
\end{equation}
where $P\left(3/2,y^{*}\right)$ is the lower incomplete Gamma ratio.
The inverse of the lower incomplete Gamma ratio can be computed numerically and, by substituting $y^{*}$ back with $m^{*}/\overline{m}_{p}(r)$ we obtain
\begin{equation}\label{massval}
 m(r)=\overline{m}_{p}inv\left(P\left(3/2,Y)\right)\right)
\end{equation}
We can therefore assign to each planetesimal its own mass value through a second Montecarlo extraction.\\
% The number of massless particles we used in our simulations has been $n_{mp}=8\times10^{4}$ to optimise the computational load.
The use of massless particles assures the linearity of the processes investigated over the number of considered bodies: we can therefore extrapolate the number of impacts expected in a disk of planetesimals described by the density profiles previously presented by multiplying the number of impacts recorded in our simulations by a factor $\gamma$ where
\begin{equation}
 \gamma=N_{tot}/n_{mp}
\end{equation}
where $N_{tot}$ is computed through Eq. \ref{ntot} and $n_{mp}=8\times10^{4}$. Note that the $\gamma$ factor depends only on the radial extension of the considered region (i.e. $r_{min}$ and $r_{max}$): we took advantage of this fact in parallelising the algorithm as described in Sect. \ref{parallel}.\\
% but does not depend explicitly from the heliocentric distance $r$ since the $n_{mp}$ planetesimals have been distributed following the number density profile $n^{*}(r)$ of the disk.\\
The dynamical evolution of Jupiter and the swarm of massless particles is computed through a fourth order Runge-Kutta integrator with a self adjusting time-step. To time-step is chosen by evaluating at each given time the smallest timescale $\tau_{min}$ between:
\begin{itemize}
 \item the orbital periods of the massless particles
 \item the orbital periods of Jupiter and the two asteroids
 \item the free-fall time of Jupiter-particle pairs considered as isolated systems
\end{itemize}
The time-step is then computed as
\begin{equation}
 t_{ts}=\tau_{min}/f_{ts}
\end{equation}
where $f_{ts}=100$ in our simulations.\\
%
% \subsection{Chemical and physical characterisation of planetesimals}
%
Finally, to estimate the amount of volatiles delivered to the two asteroids Vesta and Ceres by the planetesimals, we assumed that planetesimals formed at their initial positions and considered two compositional classes:
\begin{itemize}
 \item planetesimals formed inside the Snow Line ({\bf ISL} in the following) were considered rocky bodies;
 \item planetesimals formed beyond the Snow Line ({\bf BSL} in the following) were considered volatile-rich bodies.
\end{itemize}
ISL and BSL planetesimals were characterised by mean density values respectively of $\rho_{ISL}=3.0\,g/cm^{3}$ and $\rho_{BSL}=1.0\,g/cm^{3}$. The location of the Snow Line in our simulations was placed at $r_{SL}=4.0\,AU$ (see \cite{en08} and references therein).

\subsection{Collisional history}\label{impacts}

To reproduce the collisional histories of the two asteroids we opted for a statistical approach based on the probability density distributions of the asteroids along their orbits. Our method is similar to the analytical method developed by \cite{opi76}, yet the latter (as well as its variants) compute an average impact probability by assuming that the longitude of node, the anomaly and the argument of pericenter vary uniformly over all possible orientations. Moreover, Opik's method may fail evaluating the impact probability for near-tangent orbits and for very eccentric orbits \citep{gcv88}. Given the wide range of possible impact geometries due to the extension ($2-10$ AU) of the disk and that Jupiter's perturbations may significantly change the orbits of the planetesimals on timescales analogous to their precession timescales, we preferred the use of a numerical algorithm. This way we were able to characterise in a semi-deterministic way the impact probabilities of the real orbital configurations.
\\
Each asteroid was spread on a torus representing its spatial probability density and characterised by a mean radius $R_T$ and a section $\sigma_{T}$ defined as
\begin{equation}
 R_{T} = a_{A}
\end{equation}
and
\begin{equation}
 \sigma_{T}= \frac{\pi}{4} \times \left( D_{A} f_{G} \right)^{2}
\end{equation}
where $a_{A}$ and $D_{A}$ were respectively the semimajor axis and the physical diameter of the considered asteroid while $f_G$ is the gravitational focusing factor
\begin{equation}
 f_{G}=1+\left( \frac{v_{esc}}{v_{enc}}\right)^{2}
\end{equation}
with $v_{esc}$ being the escape velocity from the asteroid and $v_{enc}$ the relative velocity between the asteroid and the planetesimal. The gravitational focussing factor $f_{G}$ was introduced to account for the perturbations of the asteroids on the orbits of approaching planetesimals, which were not included in the explicit dynamical model.\\
When a planetesimal crosses one of the two tori, the impact probability is the probability that both the planetesimal and the asteroid will occupy the same spatial region at the same time. This probability can be evaluated as the ratio between the \emph{effective collisional time} and the orbital period of the asteroid. The effective collisional time is the amount of time available for collisions and is evaluated as the minimum between the time spent by the asteroid and the planetesimal into the crossed region of the torus. This is equivalent to writing
\begin{equation}
P_{coll}=\frac{min(\tau_{P},\tau_{A})}{T_{A}}
\end{equation}
where $T_{A}$ is the orbital period of the asteroid, $\tau_{A}$ and $\tau_{P}$ are respectively the time spent by the asteroid and the planetesimal into the crossed region of the torus while $min(\tau_{P},\tau_{A})$ is the effective collisional time.\\
To estimate $\tau_{P}$ and $\tau_{A}$ we need to identify the intersections between the orbit of the planetesimal and the torus of the asteroid. Assuming that the orbital path of the planetesimal during the timestep when it approaches the torus is linear, we just need to solve a ray--torus intersection problem.\\
A torus centred on the origin of the axes and lying on the $xy$ plane is described by the equation
\begin{equation}
 f(x,y,z)=\left( x^{2}+y^{2}+z^{2} - (R^{2}+s^{2})\right)+4R^{2}\left( z^{2}-s^{2}\right)
\end{equation}
where $R=R_{T}$ and $s=0.5\left(D_{A}f_{G}\right)$. The surface of such torus is described by the homogeneous equation $f(x,y,z)=0$. By substituting the variables $(x,y,z)$ with the vectorial equation of the ray associated to the linearised orbit
\begin{equation}
 \vec{r}=\vec{m}t+\vec{q}
\end{equation}
where $m$ and $q$ are respectively the angular coefficient and the origin of the ray, we obtain the fourth order equation identifying the intersections between the ray and the torus. By writing
\begin{equation}
 a=\vec{m}\cdot\vec{m}
\end{equation}
\begin{equation}
 b=\vec{m}\cdot\vec{q}
\end{equation}
\begin{equation}
 c=\vec{q}\cdot\vec{q}
\end{equation}
we can express the fourth order equation as
\begin{equation}
 d_{4}t^{4}+d_{3}t^{3}+d_{2}t^{2}+d_{1}t+d_{0}=0
\end{equation}
where
\begin{align}
 d_{4}=&\, a^{2} \nonumber\\
 d_{3}=&\, 4ab \nonumber\\
 d_{2}=&\, 4b^{2}-2as^{2}-2aR^{2}+2ac+4R^{2}m^{2}_{z} \\
 d_{1}=&\, 4bc-4bs^{2}-4bR^{2}+8m_{z}q_{z}R^{2} \nonumber\\
 d_{0}=&\, c^{2}+R^{4}+s^{4}-2R^{2}s^{2}-2cs^{2}-2cR^{2}+4R^{2}q^{2}_{z}. \nonumber
\end{align}
This fourth order equation can be solved analytically through Ferrari's method or numerically. The quartic equation will have up to four real solutions: by selecting the appropriate pair of real solutions, if existing, we can derive the coordinates of the intersection points $\vec{i}_{1}$ and  $\vec{i}_{2}$.\\
From the intersection points we can derive the length of the planetesimal's path $d_{P}$ through the torus and the crossing time $\tau_{P}$ which, in the linear approximation, is
\begin{equation}
 \tau_{P}=\frac{d_{P}}{|\vec{v_{P}}|}
\end{equation}
where $|\vec{v_{P}}|$ is the modulus of the velocity of the planetesimal since the velocity and the path are parallel.\\
From the intersection points we can also derive the angular width of the crossed section of the torus $\Delta\theta_{A}$ and the time $\tau_{A}$ the asteroid spends into the crossed region:
\begin{equation}
\tau_{A}=\Delta\theta_{A}\frac{T_{A}}{2 \pi}=\frac{\Delta\theta_{A}}{\omega_{A}}
\end{equation}
where $\omega_{A}=n_{A}=\frac{2\pi}{T_{A}}$ is the angular velocity (coinciding with the orbital mean motion for circular orbits) of the asteroid.\\
%
% \subsection{Impact cratering}
%
To evaluate the effects of the impacts on the two asteroids, we computed the average crater diameter produced by each collision using the empirical scaling law (see p. $165$ of \cite{dpl01})
\begin{equation}\label{craterlaw}
 D=1.8\rho_{i}^{0.11}\rho_{a}^{-1/3}\left(2R_{i}\right)^{0.13}\left(\frac{E_{k}}{g_{a}}\right)^{0.22}\left(sin\,\theta\right)^{1/3}
\end{equation}
where the indexes $a$ and $i$ indicate the asteroid and the impactor respectively, $\rho$ is the density, $R$ the physical radius, $g$ the gravitational acceleration, $E_{k}$ is the impact kinetic energy, $\theta$ is the impact angle respect to the local horizontal and all quantities are evaluated in mks units. In our first estimation of the surface cratering of the two asteroids we used a fixed value $\theta=45^{\circ}$ as the average impact angle (see \cite{pm00} and references therein).

\subsection{Parallelising the model}\label{parallel}

Before concluding the section devoted to the dynamical and physical model, we would like to briefly discuss one advantage of our modeling, i.e. the possibility to parallelise the treatment of the evolution of this template of the Solar System in a straightforward way.\\
Since the disk of planetesimals is reproduced by massless particles, which by definition do not interact between themselves, and since Jupiter's evolution is governed by the migration and accretion rates described by Eqs. \ref{coregrowth}, \ref{gasgrowth} and \ref{radiallaw}, we can apply a \emph{data parallel} approach and split each simulation into a set of sub-problems which can be treated in parallel. In our implementation, we divided the disk of planetesimals into a number of concentric rings containing a fixed amount of test particles. Each ring has been evolved independently under the influence of the forming Jupiter and the collisions with the two asteroids have been recorded. At the end of the set of simulations, we merged and temporally reordered their outputs to obtain a representation of the evolution of the system as a whole. Obviously, we had to use a different  $\gamma$ factor for each ring of massless particles to normalise it to the real population expected in the orbital region considered. However, as anticipated in Sect. \ref{disk} we can do this in a straightforward way since $\gamma$ depends only on the radial extension of the ring.\\ Through this approach, we were able to run the equivalent of a $6$-month long simulation with $8\times10^{4}$ massless particles for each migration scenario by running a set of $8$ sub-simulations with $10^{4}$ massless particles each, every sub-simulation taking about $21$ days to conclude. This duration of the sub-simulations is the reason why we concentrated on the perturbing role of Jupiter in this first investigation of the evolution of the early Solar System. The inclusion of Saturn as a second massive perturber, in fact, would have implied the doubling of the time required to complete each simulation and the same hold true for each planetary embryo considered.

\section{Results}\label{results}

In this section we will describe the results of our simulations, highlighting the differences we found between the four dynamical scenarios for the migration of Jupiter and the influence of the primordial size distribution of the planetesimals. We will first (Sect. \ref{dynamical_features}) concentrate on detailing the dynamical classes of impactors that dominate the different phases of the evolution of our template of the early Solar System, then (Sect. \ref{flux_features}) we will characterise the impact fluxes recorded for Vesta and Ceres in the scenario of planetesimal formation in a quiescent disk and discuss their effects on the two asteroids (\ref{impact_features}). Finally, we will discuss the influences of the migration timescale of Jupiter (Sect. \ref{slowmigration}) and of the scenario of planetesimal formation in a turbulent disk (Sect. \ref{turbulence}) on our results.
% In our simulations we identified $4$ distinct dynamical families contributing to the cratering of Vesta and Ceres in the temporal framework considered. Two of these families (a resonant and a non-resonant one) belonged to the ISL population and two (again, a resonant and a non-resonant one) to the BSL population.

\subsection{Dynamical features of the impactors}\label{dynamical_features}

\begin{figure*}
 \centering
 \includegraphics[width=17.5cm]{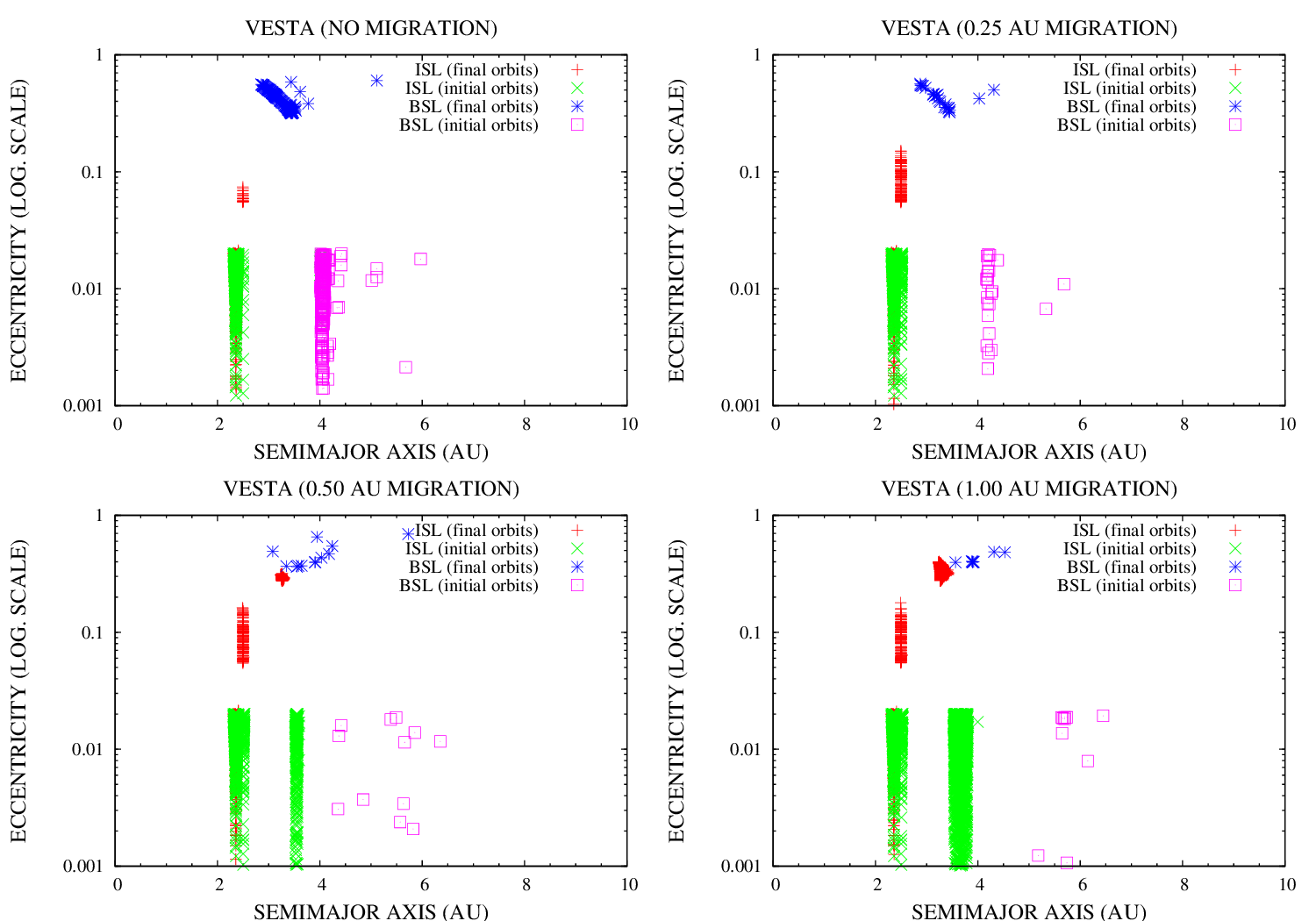}
 \caption{Semilogarithmic plots of the orbital elements of the impactors on Vesta in the $a-e$ plane. Red and green symbols represent respectively the final (at impact) and initial orbits of ISL impactors, while the blue and magenta symbols represent respectively the final and initial orbits of BSL impactors. As is clearly visible from the plots, both ISL and BSL impactors can be divided in two dynamical families: a first  non-resonant one (i.e. ISL primordial impactors and BSL scattered impactors) and a second resonant one (i.e. ISL and BSL resonant impactors). Note that these plots show only the dynamical classes of impactors recorded in the simulations: they are not normalised to the real disk population.}\label{fig1}
\end{figure*}
\begin{figure*}
 \centering
 \includegraphics[width=17.5cm]{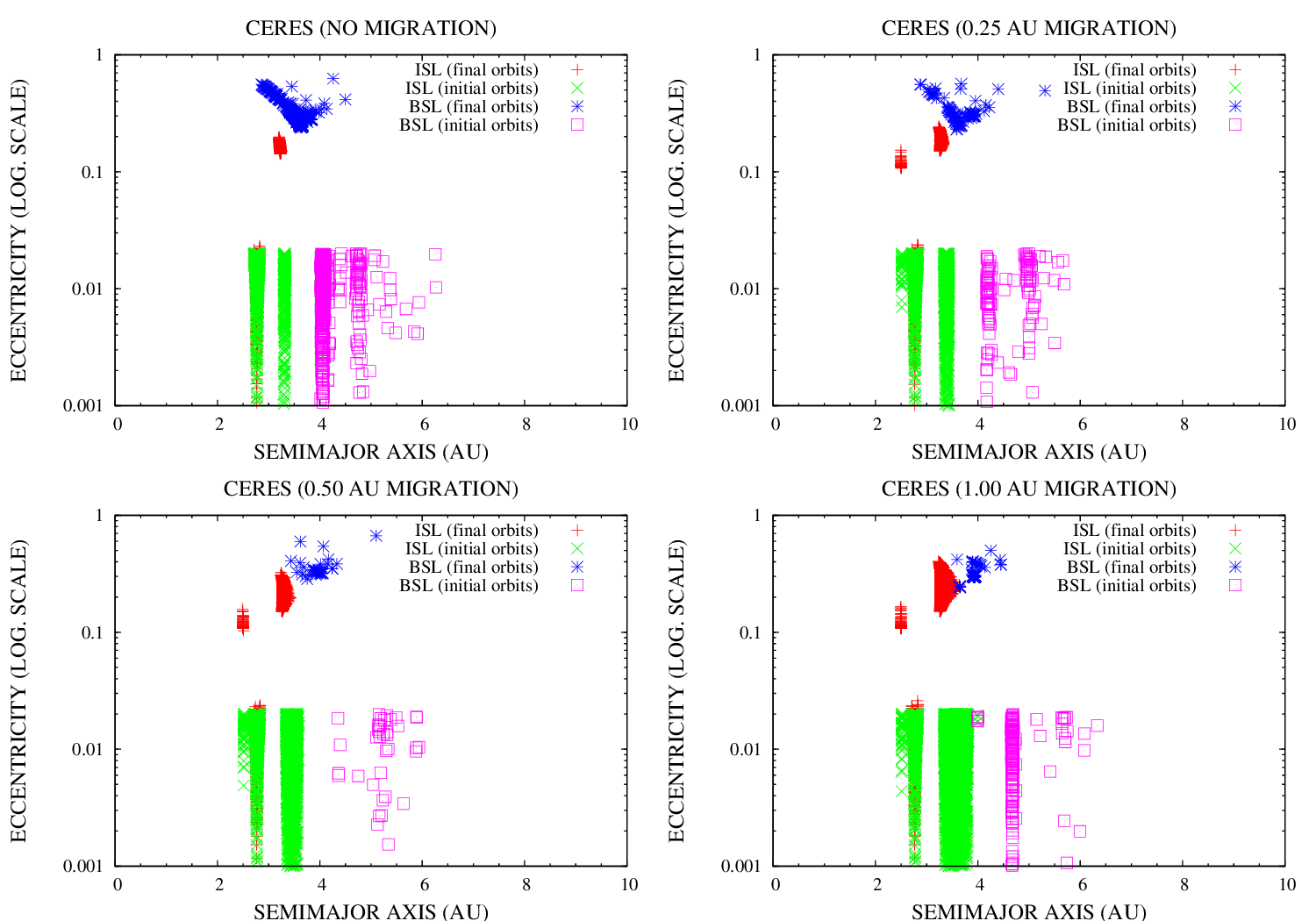}
 \caption{Semilogarithmic plot of the orbital elements of impactors on Ceres in the $a-e$ plane. Red and green symbols represent respectively the final (at impact) and initial orbits of ISL impactors, while the blue and magenta symbols represent respectively the final and initial orbits of BSL impactors. As in Fig. \ref{fig1}, both ISL and BSL impactors can be divided into a non-resonant (i.e. ISL primordial impactors and BSL scattered impactors) and a resonant (i.e. ISL and BSL resonant impactors) population. Note that these plots show only the dynamical classes of impactors recorded in the simulations: they are not normalised to the real disk population.}\label{fig2}
\end{figure*}

In our simulations we identified $2$ distinct dynamical classes (i.e. resonant and non-resonant impactores) contributing to the cratering of Vesta and Ceres for each of the populations of planetesimals (ISL and BSL) we considered (see Figs. \ref{fig1} and \ref{fig2}).\\
The non-resonant class of impactors for the ISL population of planetesimals is composed by bodies spatially near each of the two asteroids: during the first $10^{6}$ years of our simulations, these impactors are characterised by the same dynamical features they possessed at the beginning of the simulations and can be seen as the tail of the accretion process of the two asteroids. Hereafter we indicate this population as \emph {primordial impactors}. The collisions due to primordial impactors would stop once the region of space near each of the two asteroids is depleted, if not for the increasing gravitational perturbations of Jupiter that inject new massless particles on Vesta-crossing or Ceres-crossing orbits.\\
The second ISL group is composed of families of \emph {resonant impactors}. The number of active resonances and their relative abundances of impactors strongly depend on the extent of Jupiter's displacement and the target asteroid considered. We also stress that the active resonances in our simulations reflect the orbital configuration of our template of the early Solar System: the inclusion of Saturn, and to a lesser extent that of planetary embryos, would introduce new resonances in the system (see e.g. \cite{omb07}). The flux of impactors due to the resonances with Jupiter increases with increasing displacements of the giant planet (see Figs. \ref{fig1} and \ref{fig2}), as it is expected due to the sweeping of the resonances through wider spatial regions. The resonances affecting Vesta and Ceres in the temporal framework of our simulations are the $3:1$ and the $2:1$ resonances (see Figs. \ref{fig1} and \ref{fig2}): we identified them using their locations, i.e. about $2.5$ AU and $3.3$ AU, in the simulations were Jupiter forms at its present position. Resonant impactors appear in the simulations about $10^{5}$ years after the beginning of the rapid gas accretion phase, when Jupiter's mass is high enough to strongly excite the resonances (see Figs. \ref{fig3} and \ref{fig4}). In the case of Vesta, resonant impactors from the $3:1$ resonance are present in all scenarios, while those coming from the $2:1$ resonance appears only in those simulations where Jupiter migrated by $0.5$ AU or more. The case of Ceres is the opposite: the $2:1$ resonance is always present in the four migration scenarios while the $3:1$ resonance appears only in the scenarios where Jupiter started farther away from the Sun and migrated inward.\\
Concerning the BSL population of planetesimals, the non-resonant class of impactors is populated by bodies randomly injected by Jupiter on Vesta-crossing or Ceres-crossing orbits. In the following, we will refer to this class as the \emph{scattered impactors}. Their contribution is more significant on Ceres than on Vesta due to the lower orbital excitation required to reach the outer asteroid and their abundance is inversely proportional to that of the radial displacement of Jupiter. Collisions with scattered impactors are randomly distributed across the whole timespan we considered.\\
The second class of BSL impactors is composed, similarly to the case of ISL planetesimals, by \emph{resonant impactors}. The resonances involved for the BSL planetesimals are the $3:2$ (both for Vesta and Ceres) and the $7:6$ (mainly for Ceres) resonances, which again we identified in the scenario where Jupiter formed at its present position where they are respectively located at $\approx4$ and $\approx4.75$ AU (see Figs. \ref{fig1} and \ref{fig2}). Similarly to their ISL counterparts, BSL resonant impactors from the $3:2$ resonance appear in the simulations about $10^{5}$ years after the beginning of the rapid gas accretion phase (see Figs. \ref{fig3} and \ref{fig4}). However, in the case of Ceres BSL resonant impactors from the $7:6$ resonance appear about $4\times10^{5}$ years earlier, i.e. when the core of Jupiter is massive enough to excite the resonance (see Fig. \ref{fig4}).\\
% Once normalised to the real population of planetesimals and weighted on the computed impact probabilities, the contribution of the ISL and BSL resonant impactors coming respectively from the $2:1$ and $7:6$ resonances is greatly diminished respect to that of the $3:1$ and $3:2$ resonances (see Figs. \ref{fig5} and \ref{fig6}). The contribution of the $2:1$ resonance, however, is still significant, being of the order of a few thousands impact events (see Figs. \ref{fig5} and \ref{fig6} and Tables \ref{table1} and \ref{table2}) . On the contrary, the number of BSL impactors is limited to a few impacts (see Figs. \ref{fig5} and \ref{fig6} and Tables \ref{table1} and \ref{table2}) with the only exceptions of resonant impactors in the scenario where Jupiter did not migrate (both for Vesta and Ceres) and in that where Jupite migrated by $1$ AU (Ceres only). We will further discuss the flux of impactors on Vesta and Ceres in Sect. \ref{flux_features}.\\
Concerning the dynamical features of the different classes of impactors on Vesta (see Fig. \ref{fig1}), the eccentricity values of primordial impactors are, as the name suggests, in the same range as the initial ones. Vesta's ISL resonant impactors (Fig. \ref{fig1}) have final eccentricity values distributed in the range $0.05-0.2$ for objects coming from the $3:1$ resonance and $0.2-0.5$ for those coming from the $2:1$ resonance. Concerning the BSL impactors, the scattered and resonant ones share about the same range of eccentricity values, i.e. $0.3-0.7$, but while the semimajor axes of the BSL resonant impactors concentrate between $3-3.5$ AU those of the scattered impactors are more widespread ($3-6$ AU, see Fig. \ref{fig1}).\\
As can be expected, the case of Ceres is the same as that of Vesta concerning ISL primordial impactors, i.e. the eccentricity values are the same as the initial ones (see Fig. \ref{fig2}). ISL resonant impactors on Ceres coming from the $3:1$ resonance have eccentricity values distributed between $0.1-0.2$ while those coming from the $2:1$ resonance cover the range $0.1-0.4$ (see Fig. \ref{fig2}). What we observed for Vesta concerning the BSL impactors remains valid also for Ceres: the scattered and resonant ones share about the same range of eccentricity values, i.e. $0.2-0.7$, but while the semimajor axes of the BSL resonant impactors concentrate between $3-4$ AU those of the scattered impactors are more widespread ($3.5-5.5$ AU, see Fig. \ref{fig4}).

\subsection{Characterisation of the flux of impactors}\label{flux_features}

\begin{figure*}
 \centering
 \includegraphics[width=17.5cm]{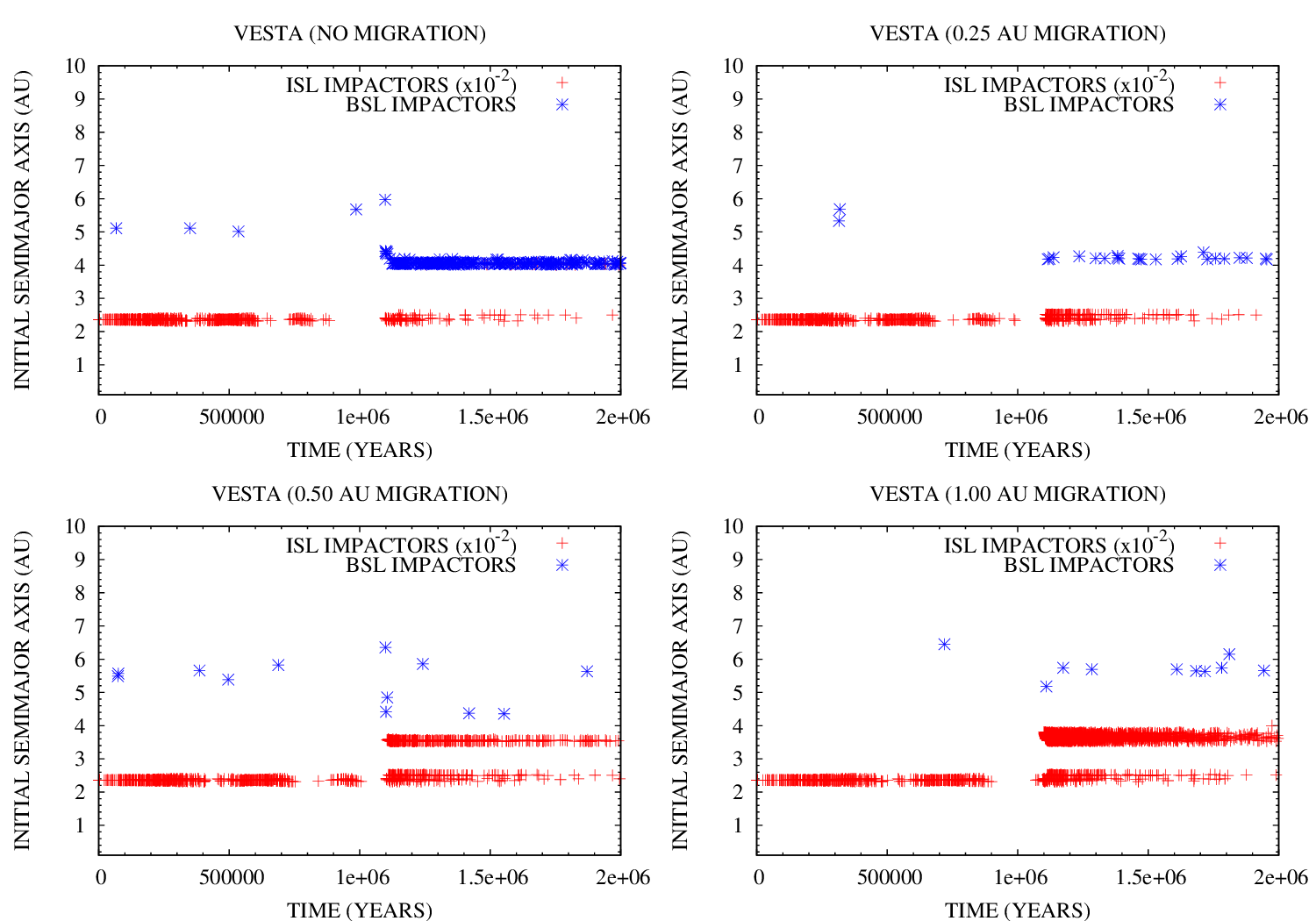}
 \caption{Temporal distribution of the impacts on Vesta due to the planetesimals showed in Fig. \ref{fig1}: the red symbols are related to ISL impactors, the blue ones to BSL impactors. ISL primordial impactors and BSL scattered impactors dominate the first $1.1\times10^{6}$ years in all scenarios, while ISL and BSL resonant impactors dominate the last $9\times10^{5}$ years. Like in Fig. \ref{fig1}, these plots show the temporal distribution of the events recorded in our simulations and they are not normalised to the real population of the disk.}\label{fig3}
\end{figure*}
\begin{figure*}
 \centering
 \includegraphics[width=17.5cm]{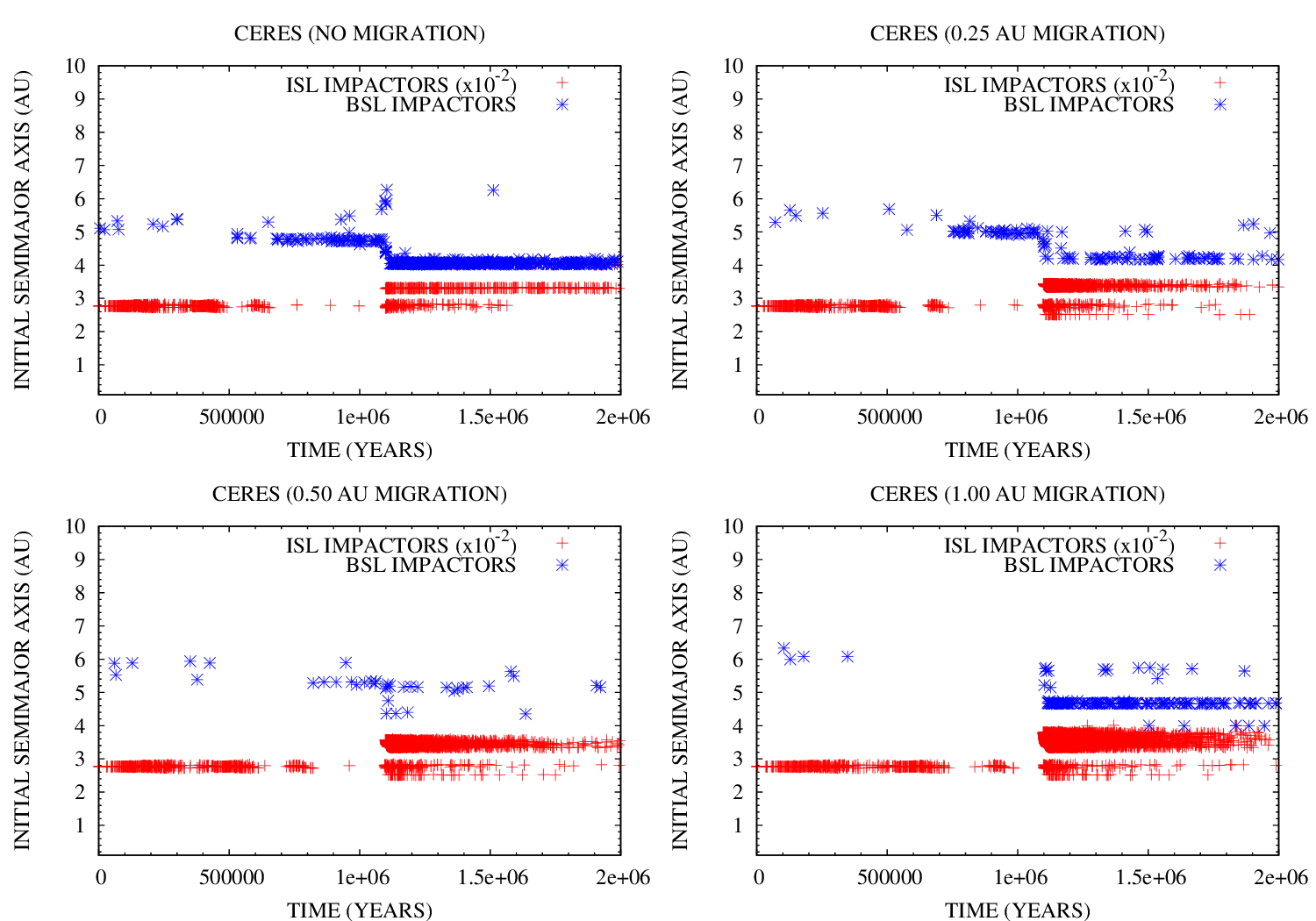}
 \caption{Temporal distribution of the impacts on Ceres due to the planetesimals showed in Fig. \ref{fig2}: the red symbols are related to ISL impactors, the blue ones to BSL impactors. Differently from the case of Vesta shown in Fig. \ref{fig3}, the first $1.1\times10^{6}$ years are dominated by the fluxes of ISL primordial impactors, BSL scattered impactors and BSL resonant impactors linked to the $7:6$ resonance. ISL and BSL resonant impactors then dominate again the last $9\times10^{5}$ years. Like in Fig. \ref{fig2}, these plots show the temporal distribution of the events recorded in our simulations and they are not normalised to the real population of the disk.}\label{fig4}
\end{figure*}
\begin{figure*}
 \centering
 \includegraphics[width=17.5cm]{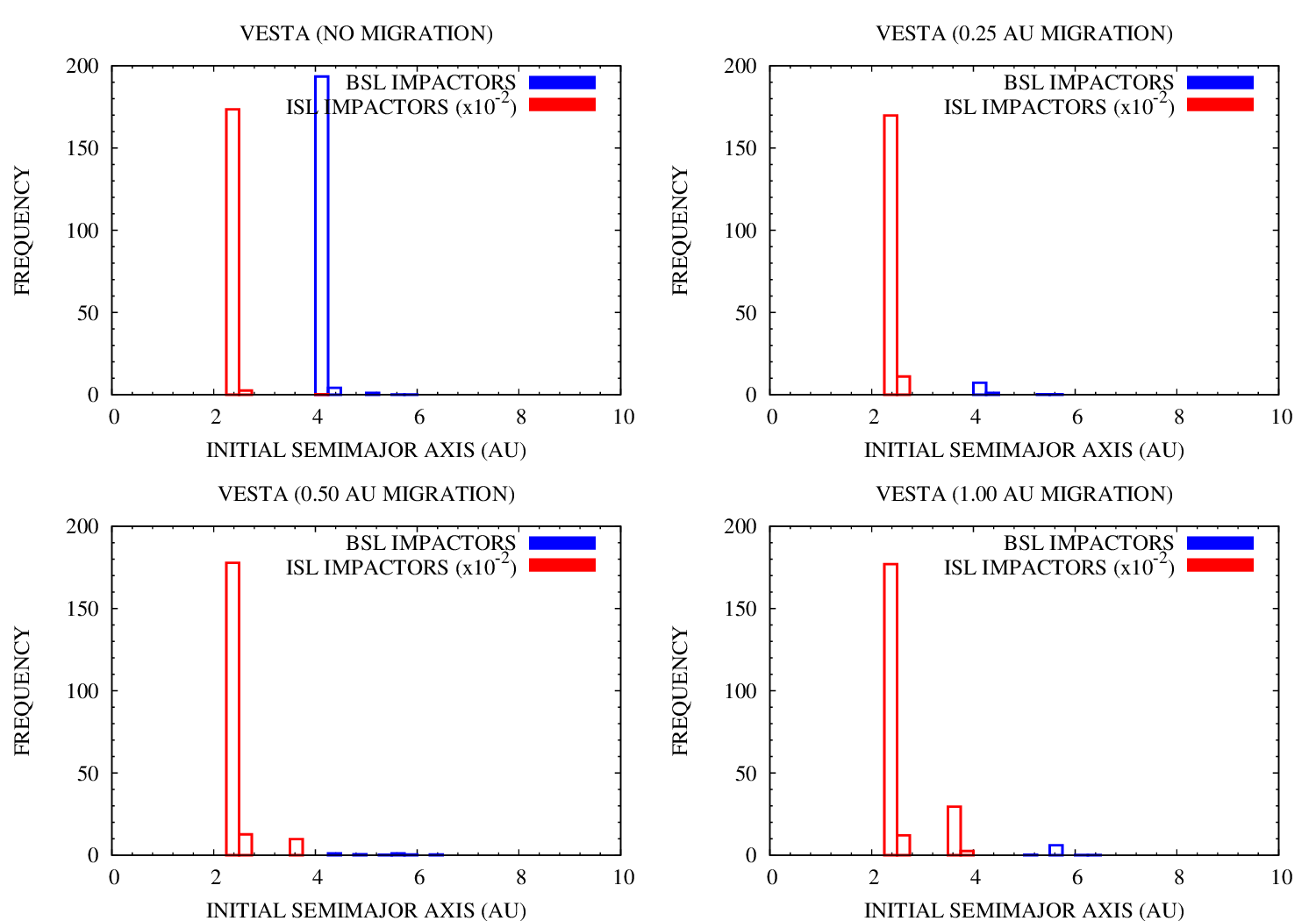}
 \caption{Frequency versus formation region histograms of the impactors on Vesta, normalised over the real planetesimal population: red bars are those related to ISL impactors, blue ones to BSL impactors. The frequency of ISL impactors in all plots has been divided by a factor $100$ to enhance the readability and facilitate the comparison of the plots. Once normalised, ISL planetesimals completely dominate the flux of impactors on Vesta. The peak in BSL impactors in the top left plot is due to the excitation of the $3:2$ orbital resonance with Jupiter.}\label{fig5}
\end{figure*}
\begin{figure*}
 \centering
 \includegraphics[width=17.5cm]{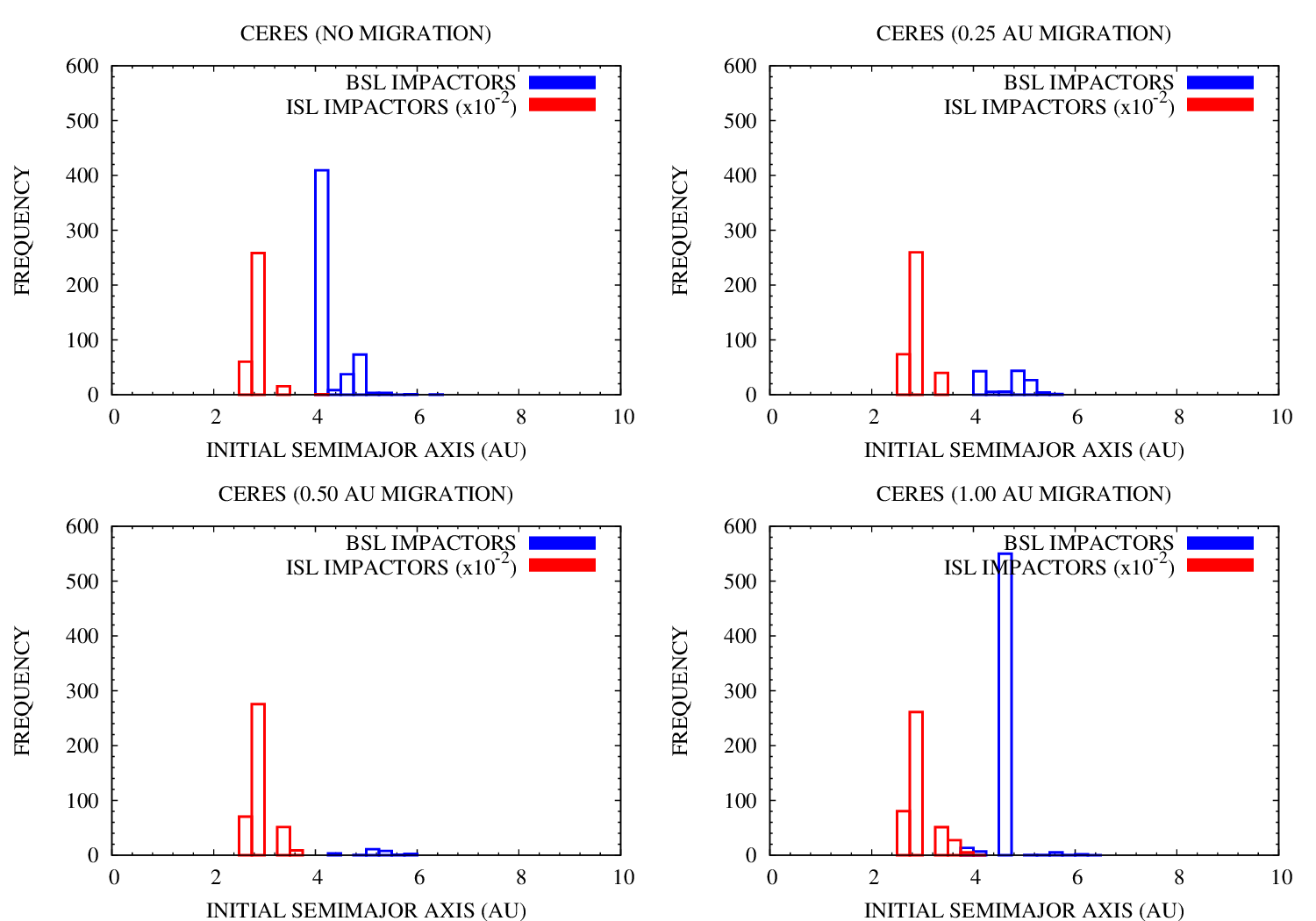}
 \caption{Frequency versus formation region histograms of the impactors on Ceres, normalised over the real planetesimal population: red bars are those related to ISL impactors, blue ones to BSL impactors. The frequency of ISL impactors in all plots has been divided by a factor $100$ to enhance the readability and facilitate the comparison of the plots. The flux of BSL impactors, while by far inferior than that of ISL bodies, is significantly higher than in the case of Vesta (see the values reported in Tables \ref{table1} and \ref{table2}). The peaks in BSL impactors are due to the excitation of the $3:2$ and, to a lesser degree, the $7:6$ orbital resonances with Jupiter.}\label{fig6}
\end{figure*}
\begin{figure*}
 \centering
 \includegraphics[width=17.5cm]{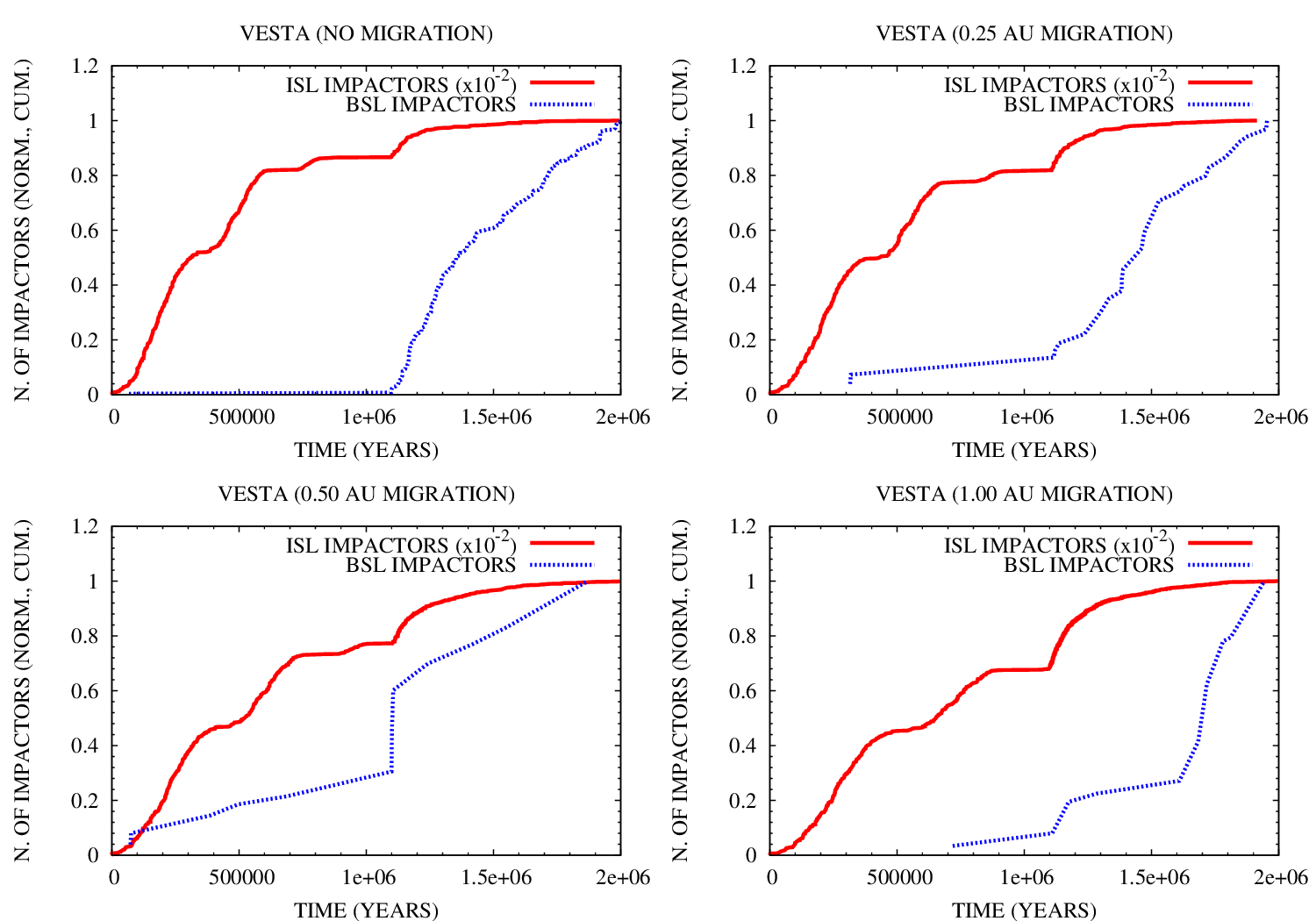}
 \caption{Normalised cumulative distribution of the flux of impactors over time for Vesta: the red curve is related to ISL impactors, the blue one to BSL impactors. In all plots the fluxes have been normalised to their values at the end of the simulations (see Table \ref{table1}). The flux of ISL impactors is divided into two main phases: a first one due to the contribution of primordial impactors and a second one, starting after Jupiter accreted a significant mass, mainly due to resonant impactors. In both phases, the cumulative distribution becomes more and more shallow towards the end, meaning that the flux eventually slows down and stops. The flux of BSL impactors begin when Jupiter's gaseous envelope reaches a significant mass and the cumulative distribution does not saturate but grows until the end of the simulations.}\label{fig7}
\end{figure*}
\begin{figure*}
 \centering
 \includegraphics[width=17.5cm]{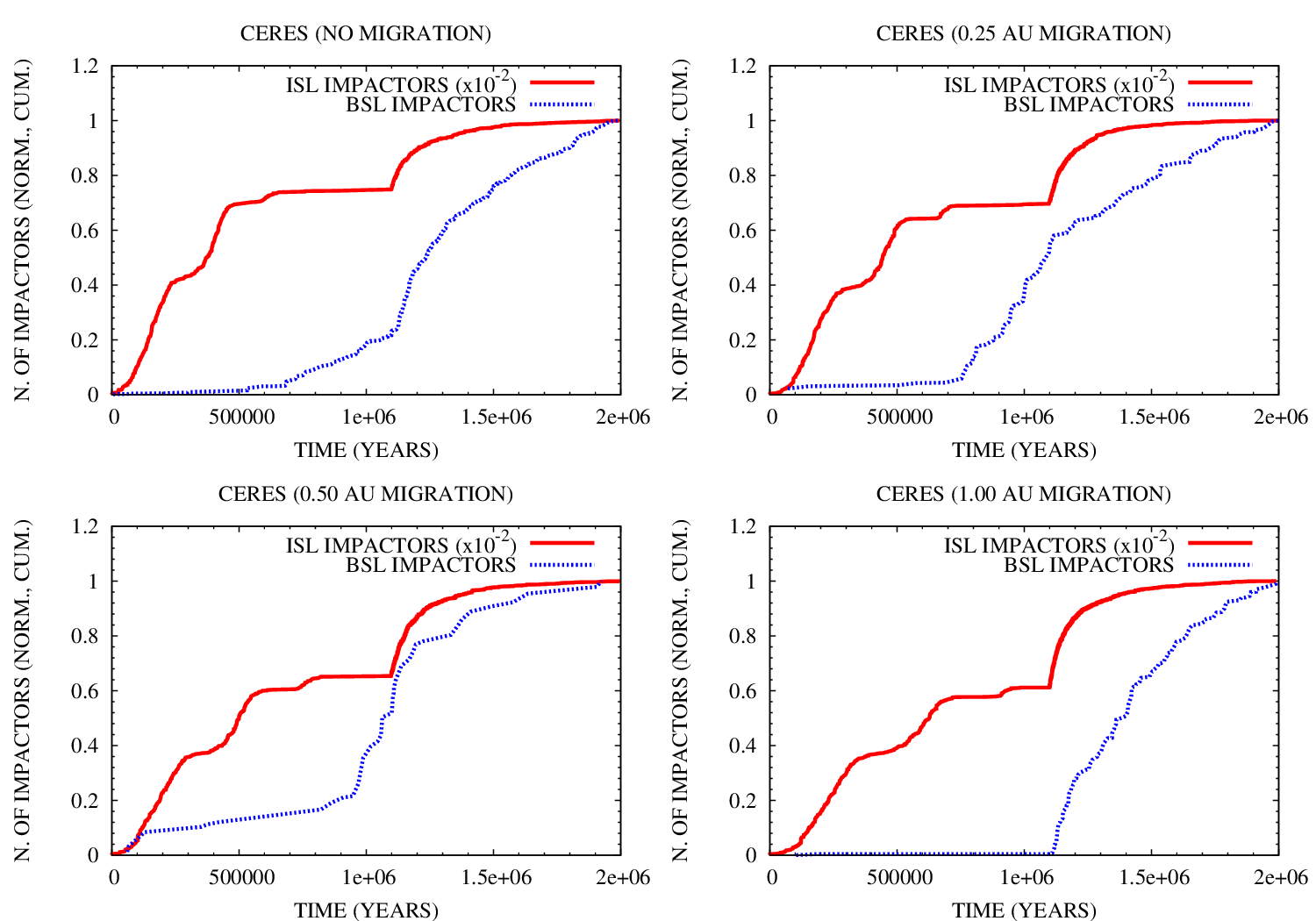}
 \caption{Normalised cumulative distribution of the flux of impactors over time for Ceres: the red curve is related to ISL impactors, the blue one to BSL impactors. In all plots the fluxes have been normalised to their values at the end of the simulations (see Table \ref{table2}). The considerations exposed for the case of Vesta (see Fig. \ref{fig7}) are valid also for Ceres. The flux of BSL impactors, however, is less erratic due to the lower orbital excitation of the planetesimals needed to reach Ceres respect to Vesta.}\label{fig8}
\end{figure*}
\begin{figure*}
 \centering
 \includegraphics[width=17.5cm]{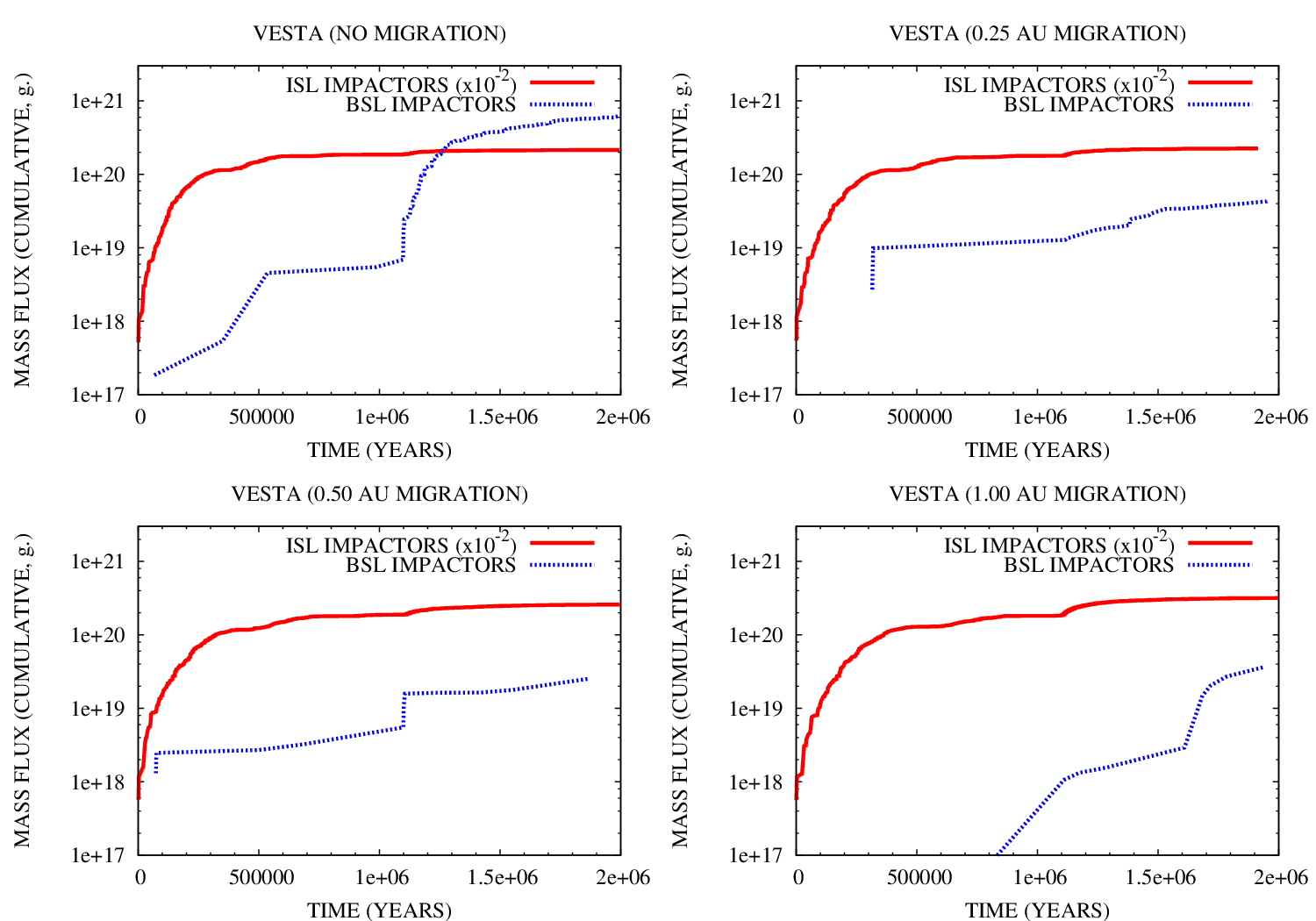}
 \caption{Cumulative distribution of the mass flux over time for Vesta: the red curve is related to ISL impactors, the blue one to BSL impactors. The mass flux of ISL impactors in all plots has been divided by a factor $100$ to enhance the readability and facilitate the comparison of the plots. The overall features of the cumulative distribution of the mass flux are similar to those of the cumulative distribution of the flux of impactors (see Fig. \ref{fig7} for details), yet these plots clearly show that the flux of BSL impactors on Vesta is erratic when Jupiter migrates, and the mass flux can consequently varies by more than one order of magnitude.}\label{fig9}
\end{figure*}
\begin{figure*}
 \centering
 \includegraphics[width=17.5cm]{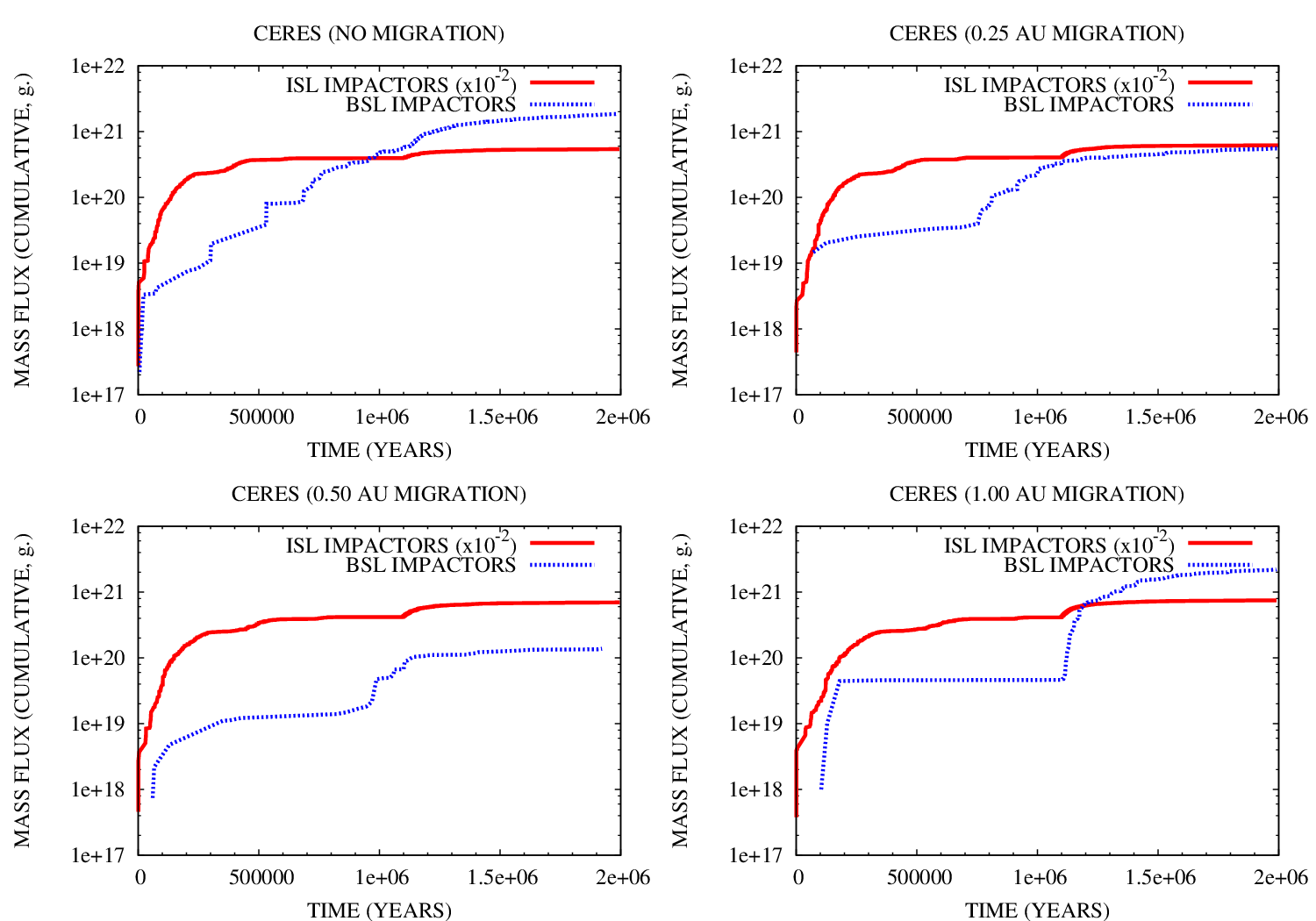}
 \caption{Cumulative distribution of the mass flux over time for Ceres: the red curve is related to ISL impactors, the blue one to BSL impactors. The number of ISL impactors in all plots has been divided by a factor $100$ to enhance the readability and facilitate the comparison of the plots. The overall features of the cumulative distribution of the mass flux are similar to those of the cumulative distribution of the flux of impactors (see Fig. \ref{fig8} for details). While the mass flux of BSL impactors varies by more than one order of magnitude between the different Jovian migration scenarios, its global features change more smoothly than in the case of Vesta.}\label{fig10}
\end{figure*}

\begin{table}
\caption{Total number of impacts on Vesta, normalised to the real population of the disk and weighted on the impact probability, for both the ISL and BSL dynamical families in the four migration scenarios of Jupiter.}
\centering
\begin{tabular}{ccc}
\hline
 \textbf{Migration scenario} & ISL impactors & BSL impactors\\
\hline
No migration & $17637$ & $199$ \\
$0.25$ AU & $18086$ & $9$ \\
$0.50$ AU & $20032$ & $4$ \\
$1.00$ AU & $22105$ & $7$ \\
\hline
\end{tabular}
\label{table1}
\end{table}

\begin{table}
\caption{Total number of impacts on Ceres, normalised to the real population of the disk and weighted on the impact probability, for both the ISL and BSL dynamical families in the four migration scenarios of Jupiter.}
\centering
\begin{tabular}{ccc}
\hline
 \textbf{Migration scenario} & ISL impactors & BSL impactors\\
\hline
No migration & $33502$ & $537$ \\
$0.25$ AU & $37355$ & $130$ \\
$0.50$ AU & $40622$ & $26$ \\
$1.00$ AU & $42522$ & $579$ \\
\hline
\end{tabular}
\label{table2}
\end{table}

The dynamical classes of impactors we showed in Figs. \ref{fig1} and \ref{fig2} and discussed in previous section represent the events recorded during our simulations and, as we anticipated, are not representative of the real flux of impactors. In Figs. \ref{fig5} and \ref{fig6} we show respectively the normalised distributions of frequency versus formation region of both ISL and BSL impactors on Vesta and Ceres in all the migration scenario we considered. The normalisation is obtained multiplying each recorded impact event for its characteristic probability, computed as described in Sect. \ref{impacts}, and the $\gamma$ factor associated to each planetesimal as defined in Sect. \ref{disk}.\\
Once normalised to the real population of planetesimals and weighted on the computed impact probabilities, the contribution on Vesta (Ceres) of the ISL and BSL resonant impactors coming respectively from the $2:1$ ($3:1$) and $7:6$ resonances is greatly diminished respect to that of the $3:1$ ($2:1$) and $3:2$ resonances (see Figs. \ref{fig5} and \ref{fig6}). The contribution of the $2:1$ ($3:1$) resonance, however, is still significant, being of the order of a few thousands impact events (see Figs. \ref{fig5} and \ref{fig6}). On the contrary, the number of BSL impactors is limited to a few impacts (see Figs. \ref{fig5} and \ref{fig6} and Tables \ref{table1} and \ref{table2}) with the only exceptions of resonant impactors in the scenario where Jupiter did not migrate (both for Vesta and Ceres) and in that where Jupiter migrated by $1$ AU (Ceres only).\\
As is clearly visible from Figs. \ref{fig3} and \ref{fig4} and Tables \ref{table1} and \ref{table2}, ISL impactors are $2-3$ orders of magnitude more abundant than BSL impactors and completely dominated the early cratering histories of the two asteroids. The frequency of BSL impactors is greatly enhanced in those migration scenarios where Jupiter excites resonances in the orbital region outside the Snow Line, as we mentioned in previous section.\\

\subsubsection{ISL impactors}\label{islbodies}

The flux of ISL impactors on the two asteroids is characterised by two phases (see Figs. \ref{fig7} and \ref{fig8}) strictly linked to the evolution of the forming Jupiter.\\
As we previously mentioned, the first phase covers the time it takes the planet to form a critical mass core and accrete a significant mass of gas, i.e. $1.1\times10^{6}$ years. During this phase the flux of impactors on the asteroids is dominated by primordial impactors. From the point of view of the number of impacts on Vesta due to ISL planetesimals, this phase accounts for $87\%-67\%$ of the total amount of collisional events (see Fig. \ref{fig7}), with a decreasing trend for increasing values of the Jovian displacement. From the point of view of the mass impacting on Vesta the contribution of this first phase varies between $86\%-57\%$ (see Fig. \ref{fig9}), again with a trend inversely proportional to the Jovian displacement. In the case of Ceres, the same ranges vary respectively between $75\%-61\%$ (see Fig. \ref{fig8}) and $73\%-55\%$ (see Fig. \ref{fig10}) with the same inverse proportionality to the Jovian displacement. The temporal distributions of the number and mass fluxes of ISL impactors during this phase are characterised by ``knees'' due to the Jovian gravitational perturbations injecting new impactors on Vesta-crossing and Ceres-crossing orbits.\\
The flux of impactors during the second phase, lasting $9\times10^{5}$ years, is instead dominated by resonant impactors originating from the $3:1$ and $2:1$ resonances, as described in Sect. \ref{dynamical_features}. As is straightforward to derive from the previous discussion, this phase accounts for $13\%-33\%$ of the number flux and $14\%-43\%$ of the mass flux on Vesta and for $25\%-39\%$ of the number flux and $27\%-45\%$ of the mass flux on Ceres. The trend across the four migration scenarios is increasing for increasing values of the Jovian displacement.\\
The total mass delivered to Vesta by ISL impactors varies between $8\%-12\%$, while for Ceres the range varies between $6\%-8\%$. The total number of impacts on Ceres is about twice as high as that on Vesta (see Tables \ref{table1} and \ref{table2}).

\subsubsection{BSL impactors}\label{bslbodies}

The flux of BSL impactors greatly varies depending on the considered asteroid and migration scenario.\\
For Vesta, the only relevant contribution of BSL impactors to the cratering history of the asteroid appears in the scenario where Jupiter did not migrate (see Figs. \ref{fig7} and \ref{fig9}). In all the other scenarios, the contribution of BSL impactors on Vesta is limited to only few events (see Table \ref{table1}). This implies that, even in the most favourable scenario, the expected flux of volatile-rich material on Vesta is about $0.23\%$ of the mass of the asteroid and is about $3\%$ of the mass delivered through rocky bodies in the same scenario.\\
For Ceres the variations between the different scenarios are somewhat less erratic, yet they span over about an order of magnitude both concerning the number and the mass fluxes (see Figs. \ref{fig8} and \ref{fig10} and Table \ref{table2}). The $3:2$ and $7:6$ resonances are the major driver of the delivery of volatile-rich material (see Fig. \ref{fig6}). Even for Ceres, however, the delivery of volatile-rich impactors is limited to about $0.24\%$ of the mass of the asteroid or about $3\%$ of the mass delivered through rocky bodies in the same scenario.\\
It must be taken into account that the amount of volatile-rich material delivered by BSL impactors and retained by each target asteroid is only a small fraction of the previously reported values. Impact velocities of BSL planetesimals are always of the order of a few km s$^{-1}$, therefore collisions will cause the erosion of the target asteroids and, due to its volatile-rich composition, the vaporisation of part or all the impactors. According to the data supplied by the Deep Impact experiment \citep{wea07}, the bulk velocity of the gas emitted by comet 9P/Tempel 1 was $\approx600$ m s$^{-1}$, while the outer regions of the outburst reached velocities of the order of $1$ km s$^{-1}$. These velocities exceed the escape velocities from Vesta and Ceres, which are respectively $\approx370$ m s$^{-1}$ and $\approx510$ m s$^{-1}$. As a consequence, volatiles vaporised by the impact will be lost to the target asteroids.
%  Ejection velocities of fragments are usually of the order of $100$ m s$^{-1}$, as suggested by theoretical models \citep{baa99}, while those of dust particles are generally lower than $400$ m s$^{-1}$ as showed by the results of the Deep Impact experiment \citep{mea05,iaa08}. Since the escape velocity from Vesta and Ceres are respectively $\approx370$ m s$^{-1}$ and $\approx510$ m s$^{-1}$, most of the excavated material will be re-captured by the asteroids.

\subsection{Characterisation of the impacts}\label{impact_features}

\begin{figure*}
 \centering
 \includegraphics[width=17.5cm]{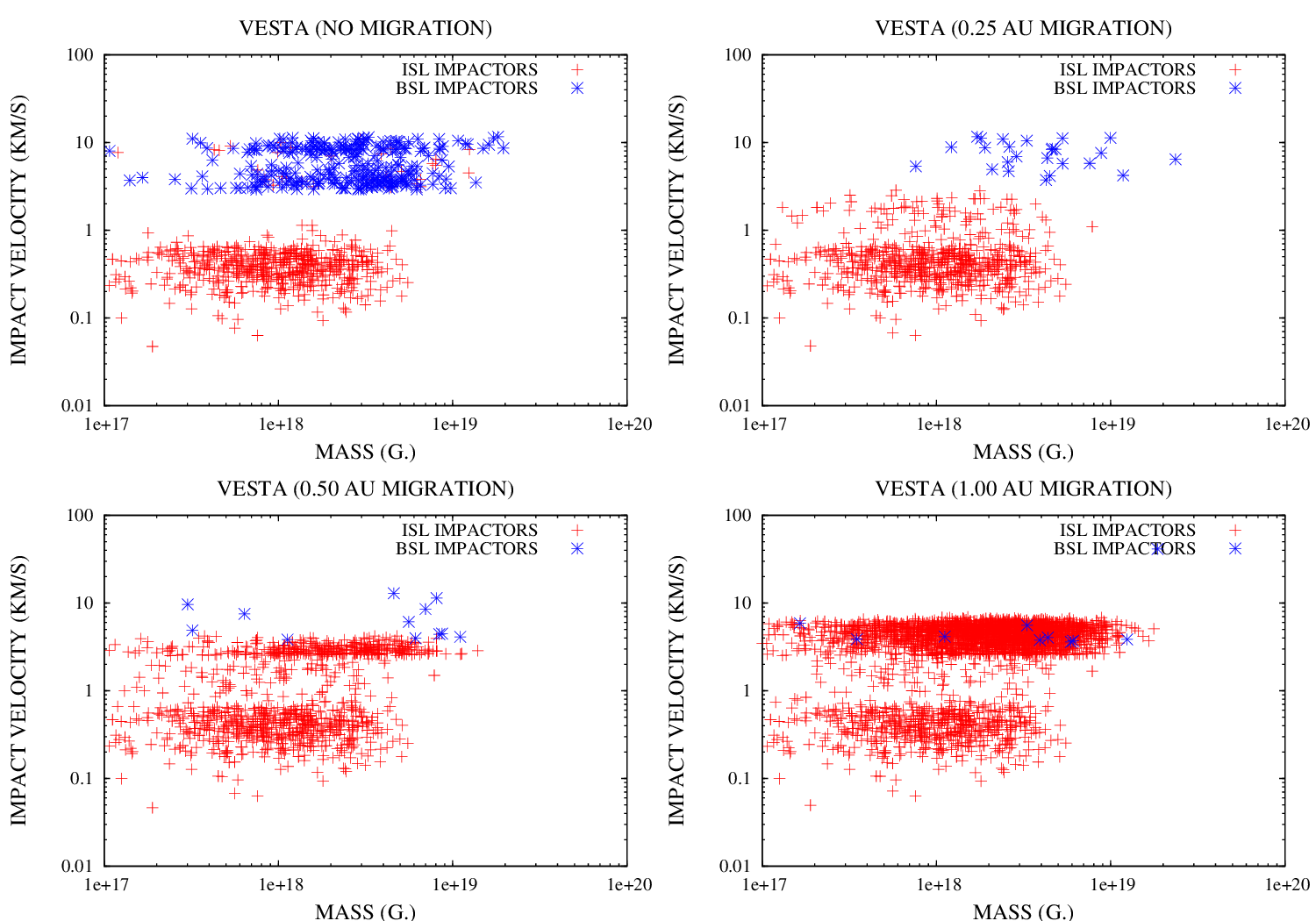}
 \caption{Dynamical characterisation of the impacts on Vesta: red symbols indicated ISL impactors and blue ones BSL impactors. The data reproduced in the plots are the events recorded in our simulations: they have not been normalised to the real population of the planetesimal disk. As can be seen from the plots, ISL resonant and primordial impactors are characterised by different dynamical features, while no difference is evident between BSL scattered and resonant impactors in those scenario with Jupiter migrating. In the no-migration scenario BSL impactors seem to divide in two clusters, a first between $3-5$ km/s and a second between $8-10$ km/s. The relative abundance of bodies in the BSL scattered and resonant groups in Fig. \ref{fig1}, however, argues against attributing this effect to the non-resonant versus resonant subdivision.}\label{fig11}
\end{figure*}
\begin{figure*}
 \centering
 \includegraphics[width=17.5cm]{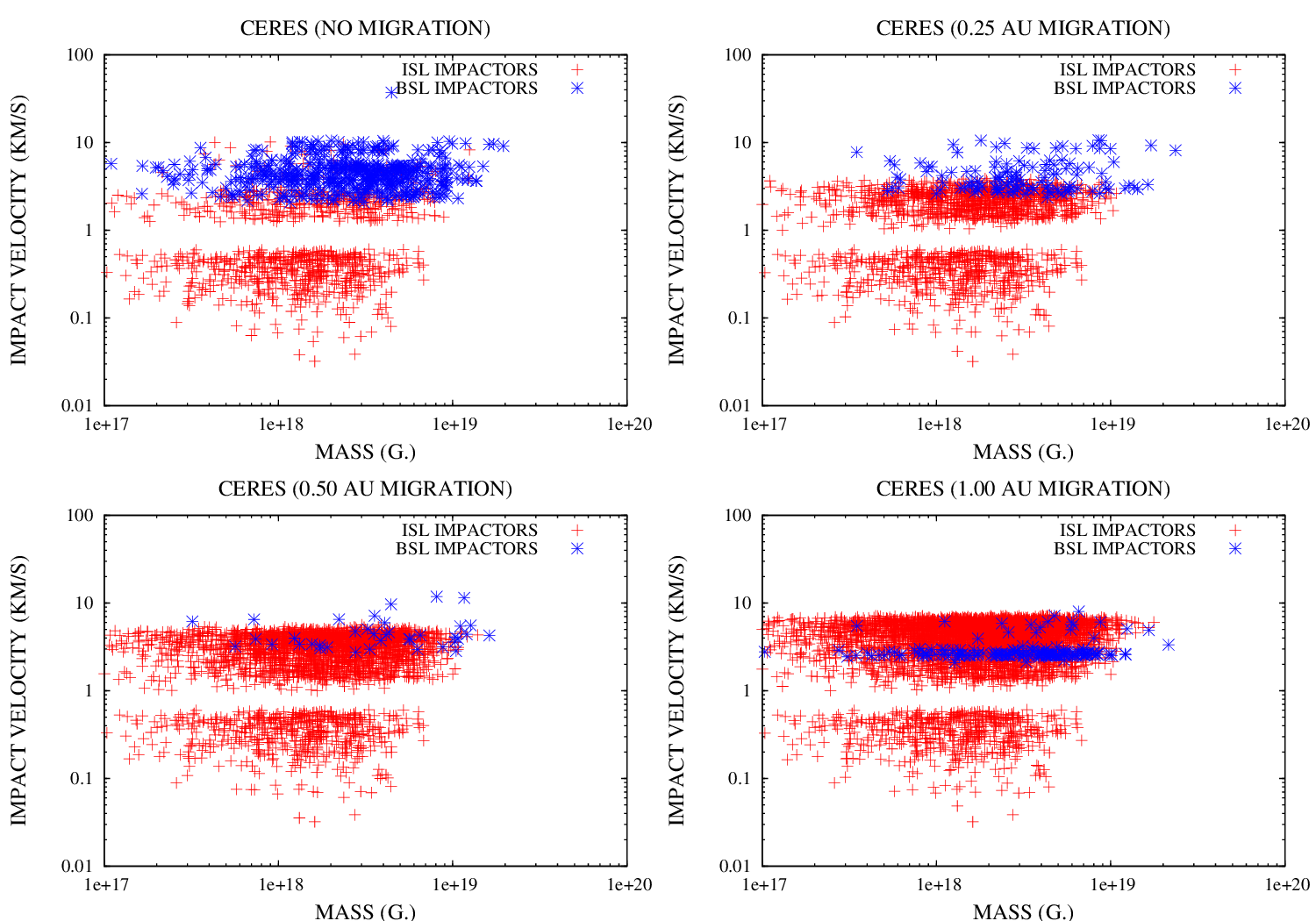}
 \caption{Dynamical characterisation of the impacts on Ceres: red symbols indicated ISL impactors and blue ones BSL impactors. The data reproduced in the plots are the events recorded in our simulations: they have not been normalised to the real population of the planetesimal disk. As can be seen from the plots, ISL resonant and primordial impactors are characterised by different dynamical features, while no difference is evident between BSL resonant and non-resonant impactors in the first three migration scenarios. In the $1$ AU migration scenario, resonant BSL impactors from the $3:2$ resonance cluster in the range $2-3$ km/s while non-resonant BSL impactors and the few resonant ones from the $7:6$ resonance are characterised by higher velocities.}\label{fig12}
\end{figure*}
\begin{figure*}
 \centering
 \includegraphics[width=17.5cm]{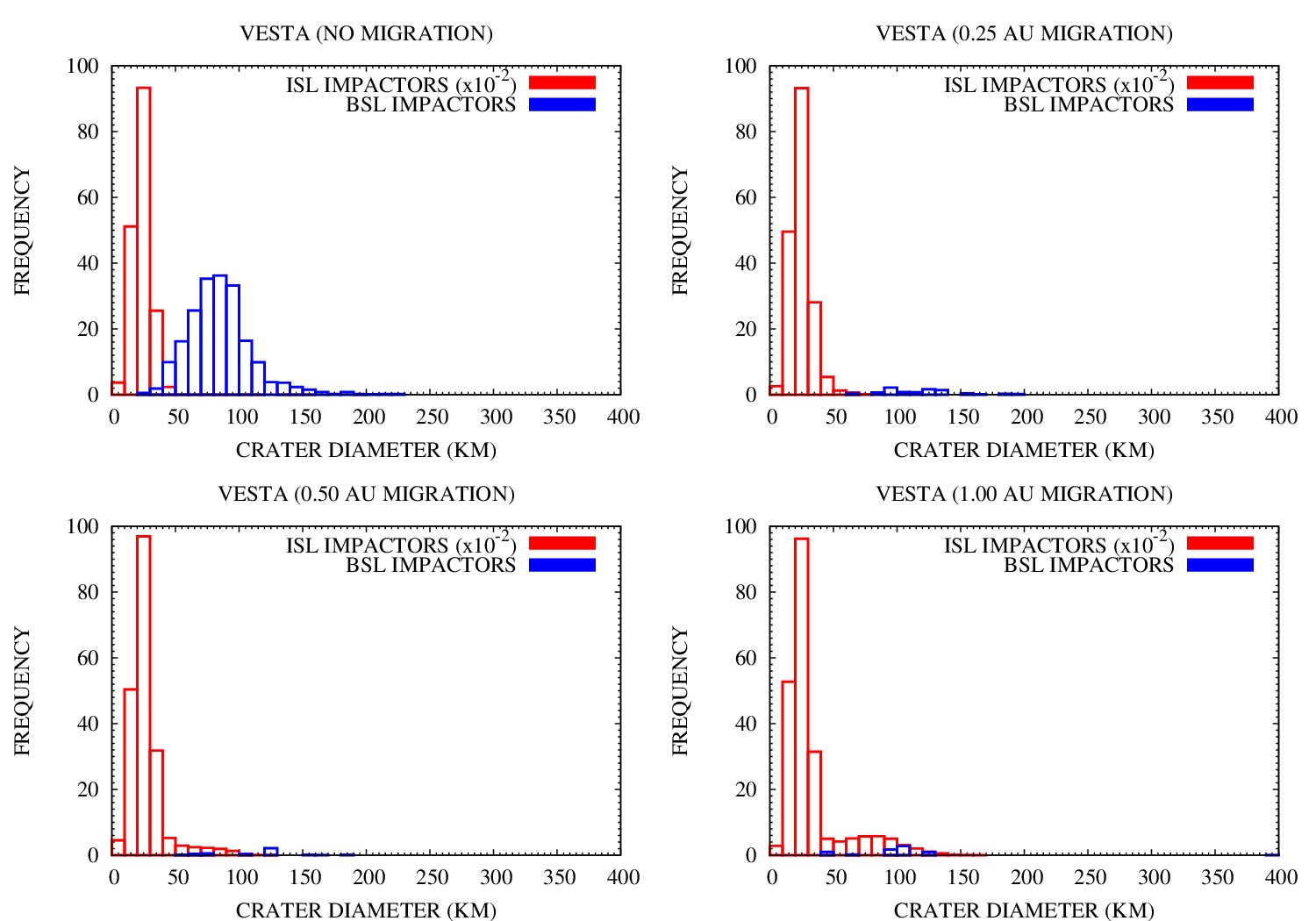}
 \caption{Normalised frequency versus crater size histograms of the impacts on Vesta: red bars are those related to ISL impactors, blue ones to BSL impactors. The frequency of ISL impactors in all plots has been divided by a factor $100$ to enhance the readability of the plots. As can be easily seen, the high-end tail of the ISL crater size distribution moves towards bigger sizes for increasing values of the Jovian radial displacement. The high-end tail of the BSL distribution shows a somewhat opposite behaviour and its contribution to the surface cratering is eventually completely masked by the one of ISL impactors, with the only exception of stochastic events producing craters bigger than $200$ km (see bottom right panel).}\label{fig13}
\end{figure*}
\begin{figure*}
 \centering
 \includegraphics[width=17.5cm]{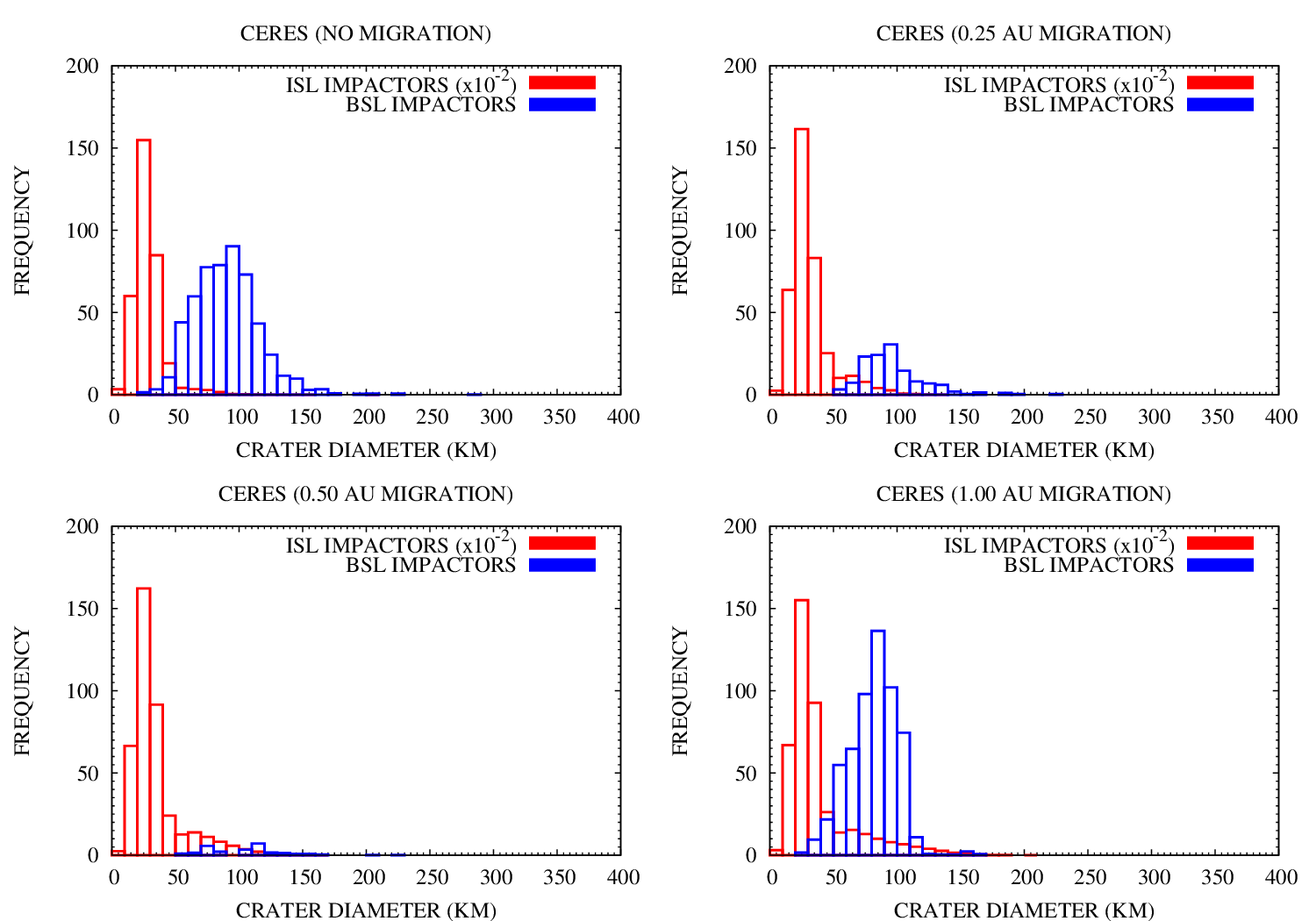}
 \caption{Normalised frequency versus crater size histograms of the impacts on Ceres: red bars are those related to ISL impactors, blue ones to BSL impactors. The frequency of ISL impactors in all plots has been divided by a factor $100$ to enhance the readability of the plots. Similarly to what we observed for Vesta in Fig. \ref{fig13}, the high-end tail of the ISL crater size distribution moves towards bigger sizes for increasing values of the Jovian radial displacement. The contribution of BSL impactors to the surface cratering is eventually masked by the one of ISL impactors.}\label{fig14}
\end{figure*}
\begin{figure*}
 \centering
 \includegraphics[width=17.5cm]{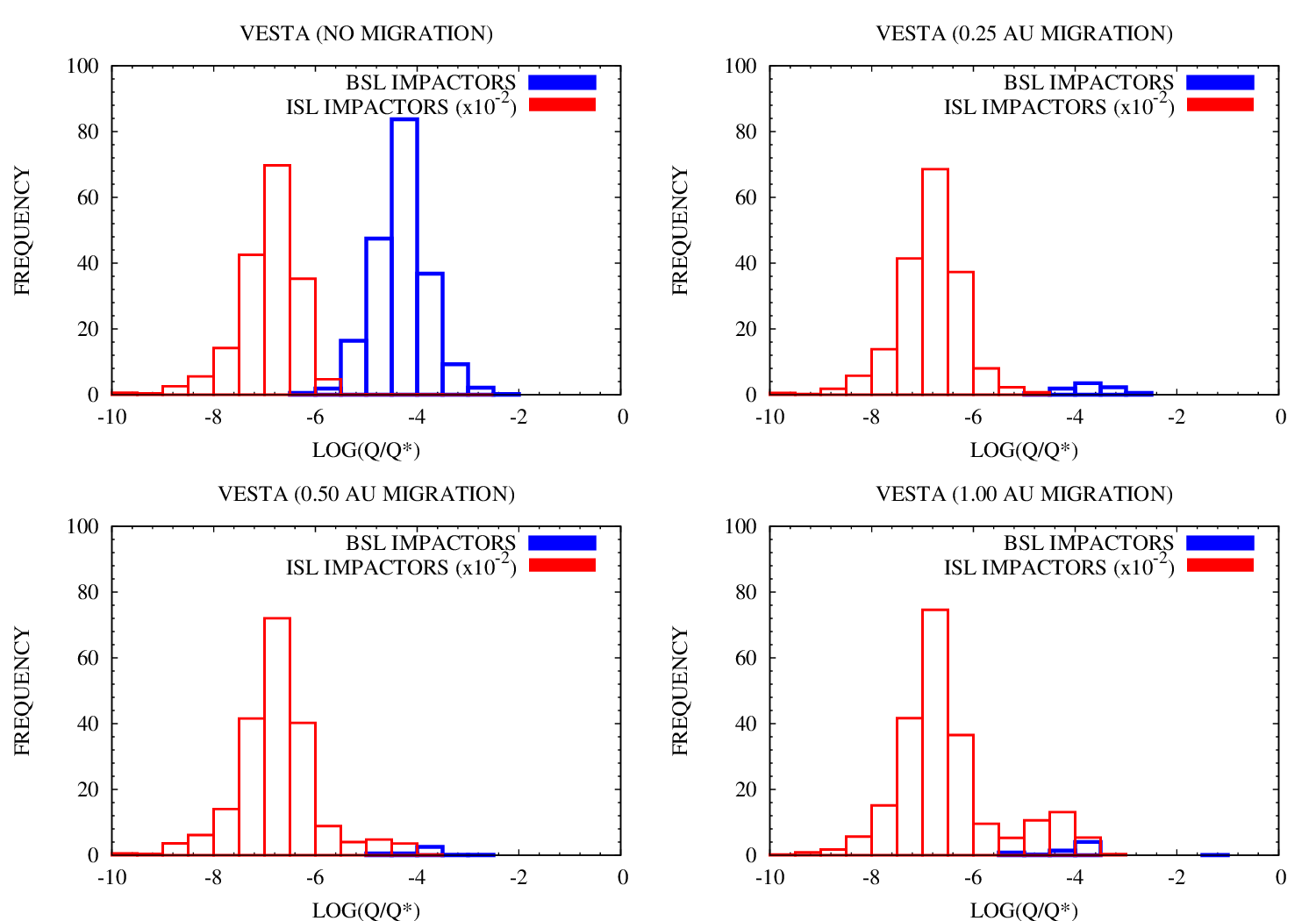}
 \caption{Normalised frequency versus impact energy (in units of the dispersal energy $Q^{*}_{D}$) histograms of the impacts on Vesta: red bars are those related to ISL impactors, blue ones to BSL impactors. The frequency of ISL impactors in all plots has been divided by a factor $100$ to enhance the readability of the plots. The most energetic impacts (of the order of $10^{-2}$) are due to BSL bodies in those scenarios where the Jovian displacement is limited (no migration or $0.25$ AU). ISL impacts in all scenarios and BSL impacts in the $0.5$ and $1$ AU migration scenarios are limited to impact energies inferior to $10^{-3}$, with the only exception of the $400$ km wide crater reported in Fig. \ref{fig13}. This event, while characterised by an low probability ($p\approx9\%$) even after normalisation, would deliver a significant fraction (about $10^{-1}$) of the energy needed to disrupt Vesta and orbitally disperse the resulting fragments.}\label{fig15}
\end{figure*}
\begin{figure*}
 \centering
 \includegraphics[width=17.5cm]{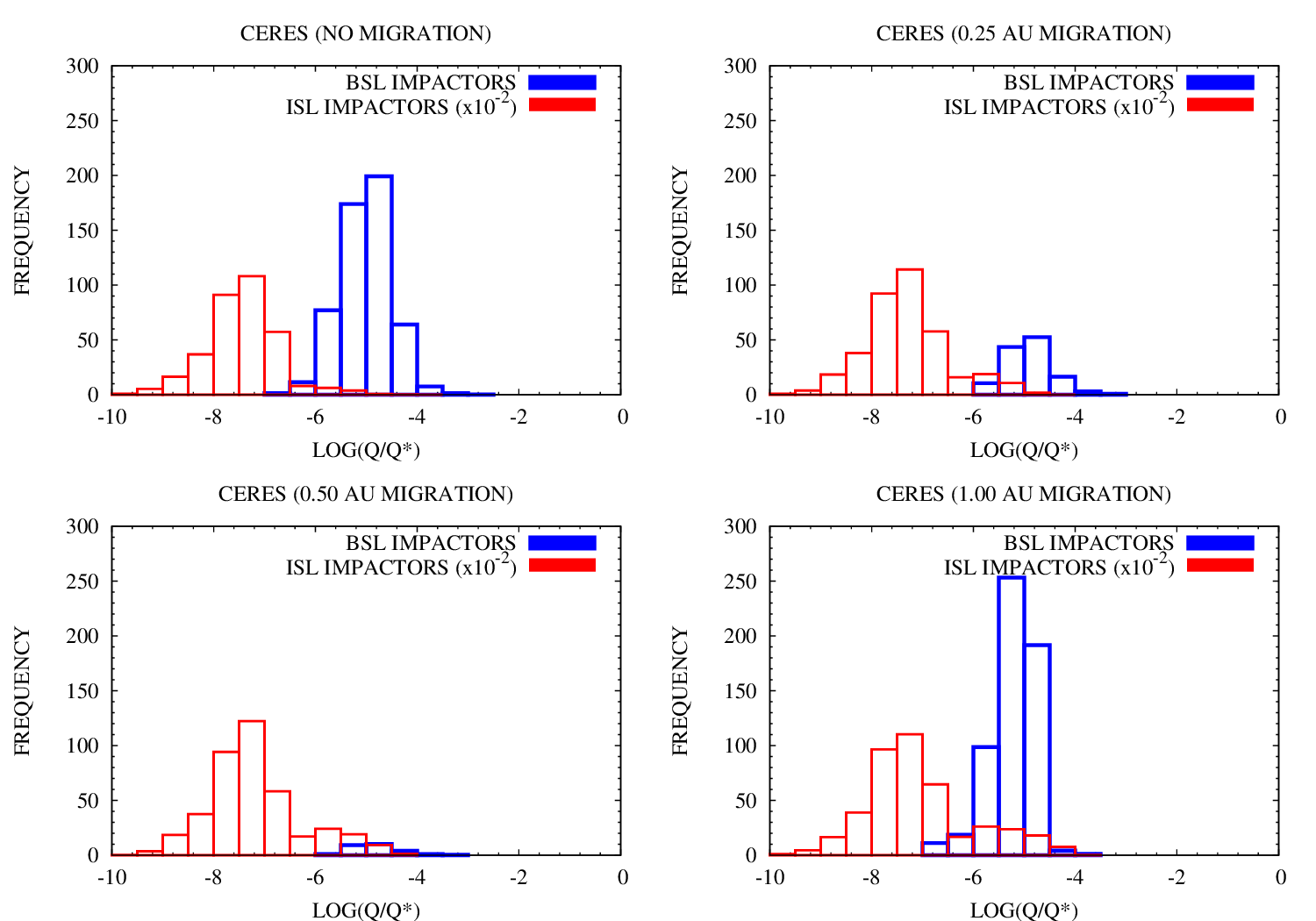}
 \caption{Normalised frequency versus impact energy (in units of the dispersal energy $Q^{*}_{D}$) histograms of the impacts on Ceres: red bars are those related to ISL impactors, blue ones to BSL impactors. The frequency of ISL impactors in all plots has been divided by a factor $100$ to enhance the readability of the plots. Due to the bigger size of Ceres, the impact energies delivered by both ISL and BSL planetesimals in all migration scenarios are of the order of $10^{-3}$ or lower.}\label{fig16}
\end{figure*}

% We recorded the characteristics of each impact in our simulations, detailing them by their dynamical features and their effects on the target (i.e. the size of the crater and an estimate of the energy delivered to the asteroid).\\
%
From the point of view of their dynamical features, the impacts we recorded in our simulations can be divided into three classes. For what it concerns the impact velocity, in fact, there is little or no difference between BSL scattered impactors and BSL resonant impactors.\\
The first class of impacts is that due to ISL primordial impactors: it is characterised by low impact velocities, always inferior to $1$ km/s for Vesta and to $600$ m/s for Ceres, and a continuous mass spectrum up to about $8\times10^{18}$ g (see Figs. \ref{fig11} and \ref{fig12}).\\
The second class is composed by impacts due to ISL resonant impactors: this class too has a continuous mass spectrum, which extends up to about $2\times10^{19}$ g (see Figs. \ref{fig11} and \ref{fig12}). The characteristic impact velocities are systematically higher than those of the first class: they range between $1-10$ km/s depending on the resonance originating the impactors (see Figs. \ref{fig11} and \ref{fig12}).\\
Finally, the third class of impacts is that due to the BSL planetesimals: as we anticipated, in fact, there is little or no evident distinction between the impacts due to BSL resonant and scattered impactors (see Figs. \ref{fig11} and \ref{fig12}). The only exception to this rule is the case of the $1$ AU migration scenario for Ceres, were BSL resonant impactors cluster between $2-3$ km/s while BSL scattered impactors are spread between $3-8$ km/s (see Fig. \ref{fig12}). In general, impact velocities of BSL impactors are spread over the range $2-10$ km/s (see Figs. \ref{fig11} and \ref{fig12}). For each of the two asteroids, however, we recorded a isolated, extreme case of about $40$ km/s (see Fig. \ref{fig11} and \ref{fig12}).\\
The mass spectrum of BSL impactors mostly overlap with that of ISL resonant impactors, extending up to about $2-3\times10^{19}$ g. (see Figs. \ref{fig11} and \ref{fig12}), yet it favours masses higher than $3\times10^{17}$ g. as a consequence of the mass-heliocentric distance relationship assumed in the model through Eq. \ref{masslaw}.\\
%
% \subsubsection{Impact cratering}
%
In the scenario where Jupiter does not migrate while forming, the craters produced by ISL and BSL planetesimals form two different populations. For increasing values of the Jovian displacement, however, the contribution of ISL resonant planetesimals tends to obliterate that of BSL planetesimals for crater diameters up to $200$ km.\\
For both asteroids, ISL planetesimals produce an asymmetrical size-frequency distribution of craters, centred at $d\approx30$ km, that dominates the low-end tail, i.e. $d\leq50$ km, of the size spectrum of the craters in all scenarios (see Figs. \ref{fig13} and \ref{fig14}). ISL resonant impactors are responsible for the asymmetry in the size-frequency distribution, causing a high-end tail that generally reaches up to $d\approx100$ km. For the highest values of the Jovian displacement we considered, this high-end tail extends up to $d\approx150-200$ km (see Figs. \ref{fig13} and \ref{fig14}) as craters produced by ISL resonant impactors grow more and more abundant as discussed in Sect. \ref{flux_features}.\\
The size-frequency distribution of craters produced by BSL planetesimals is also asymmetrical: it extends mainly between $30-200$ km and is centred about at $d\approx80$ km (see Figs. \ref{fig13} and \ref{fig14}). For Vesta, the BSL size-frequency distribution of craters becomes almost flat in all scenarios where Jupiter migrated (see Fig. \ref{fig13}). For Ceres, the BSL distribution remains bell-shaped in three scenarios over four (see Fig. \ref{fig14}). In our simulations, all craters greater than $150$ km produced on Vesta and Ceres are due to BSL planetesimals (see Figs. \ref{fig13} and \ref{fig14}): the only exception to this rule is the scenario where Jupiter migrate by $1$ AU, where also the high-end tail of the crater size spectrum is dominated by ISL impacts. It is interesting to note that, in the scenario where Jupiter migrates by $1$ AU, the bombardment of BSL planetesimals produced one cratering event with diameter $d\approx400$ km on Vesta (see Fig. \ref{fig13}). Such event, however, is characterised by an extremely low probability ($p\approx9\%$) even when normalised to the real population of planetesimals in the disk.\\
As we previously mentioned, for increasing values of the radial displacement of Jupiter the contribution of ISL resonant planetesimals grow in importance and the high-end tail of the size-frequency distribution of their craters completely overcomes the contribution of BSL planetesimals (see Figs. \ref{fig13} and \ref{fig14}).\\
%
% \subsubsection{Impact energy}
%
To give an estimate of the effects of the bombardment of ISL and BSL planetesimals on the internal structure of the two asteroids, we computed the specific impact energy $Q$ of the planetesimals expressed in units of the catastrophic disruption energy $Q^{*}_{D}$ of Vesta and Ceres (see Figs. \ref{fig15} and \ref{fig16}). The specific impact energy is defined as $Q=0.5m_{p}v^{2}_{p}/M_{T}$, where $m_{p}$, $v_{p}$ are the mass and velocity of the projectile and $M_{T}$ is the mass of the target asteroid, while the catastrophic disruption energy is defined as the specific impact energy leading to the break-up of the target asteroid with a largest fragment containing $50\%$ of its original mass. We evaluated the catastrophic disruption threshold $Q^{*}_{D}$ of Vesta and Ceres using Eq. $6$ from \citet{baa99} and the coefficients for basaltic targets computed by these authors (see Table $3$, ibid). We used the coefficients of the $v_{i}=3\,km\,s^{-1}$ case (ibid) for all impact events with a velocity lower than $3\,km\,s^{-1}$ and those of the $v_{i}=5\,km\,s^{-1}$ (ibid) for all the other impact events.\\
In the case of Vesta, the impact energy spans in the range $10^{-10}-10^{-2}$ (Fig. \ref{fig15}) yet impacts delivering more than $0.1\%$ of $Q^{*}_{D}$ are extremely limited in number. In the case of Ceres, due to its greater mass, the upper limit of the range is shifted by an order of magnitude, i.e. the energy varies between $10^{-10}-10^{-3}$ (Fig. \ref{fig16}), with very few impact events delivering more than $0.01\%$ of $Q^{*}_{D}$. The single event excavating a $d\approx400$ km-wide crater on Vesta would actually deliver about $10\%$ of the $Q^{*}_{D}$ of the asteroid. According to \citet{baa99}, such an impact would have caused the disruption of the asteroid leaving a largest fragment containing about $80\%$ of its original mass (see Fig. $10$ and Eq. $8$, ibid).\\
%
% \subsubsection{Surface excavation}
%
\begin{table*}
\caption{Collisional erosion of Vesta due to planetesimals formed in a quiescent disk in the four migration scenarios of Jupiter. N.B.: the excavated depth is estimated is all cases where Vesta is not collisionally ablated assuming that the final radius of the asteroid is the present one.}
\centering
\begin{tabular}{cccc}
\hline
 \textbf{Migration scenario} & & \textbf{Excavation depth} & \\
\hline
 & ISL impactors & ISL impactors & BSL impactors\\
 & & (post-core) & \\
\hline
No migration & $20.80$ km & $5.32$ km & $10.66$ km \\
$0.25$ AU & $23.32$ km & $8.55$ km & $1.19$ km \\
$0.50$ AU & $49.31$ km & $36.53$ km & $0.49$ km \\
$1.00$ AU & Ablation & Ablation & $0.85$ km \\
\hline
\end{tabular}
\label{table3}
\end{table*}
\begin{table*}
\caption{Collisional erosion of Ceres due to planetesimals formed in a quiescent disk in the four migration scenarios of Jupiter. N.B.: the excavated depth is estimated is all cases where Ceres is not collisionally ablated assuming that the final radius of the asteroid is the present one.}
\centering
\begin{tabular}{cccc}
\hline
 \textbf{Migration scenario} & & \textbf{Excavation depth} & \\
\hline
 & ISL impactors & ISL impactors & BSL impactors\\
 & & (post-core) & \\
\hline
No migration & $25.41$ km & $13.80$ km & $9.70$ km \\
$0.25$ AU & $42.48$ km & $31.45$ km & $2.81$ km \\
$0.50$ AU & $72.11$ km & $61.91$ km & $0.74$ km \\
$1.00$ AU & $118.20$ km & $109.60$ km & $7.19$ km \\
\hline
\end{tabular}
\label{table4}
\end{table*}
Finally, to characterise the cumulative effects of the recorded impacts on the two asteroids, we computed the total volume excavated by the Jovian bombardment under the simplifying assumption that the craters were distributed uniformly on the surfaces of Vesta and Ceres.  This approach overestimates the cratering rates at the polar regions respect to those at the equatorial regions, but it provides an interesting insight on the collisional evolution of the two asteroids. Our results for both the ISL and BSL populations of impactors are summarised in Tables \ref{table3} and \ref{table4}, where they are expressed in terms of the depth of a spherical shell whose volume is equal to that of the excavated material and whose inner radius coincides with the present radius of the relevant asteroid. Since, due to their lower impact velocities and their proximity to each asteroid, ISL primordial impactors would likely have contributed to the accretion of Vesta and Ceres more than to their cratering histories, we also estimate the contribution of the ISL resonant impactors alone by considering only ISL planetesimals impacting Vesta or Ceres after Jupiter started to accrete its gaseous envelope. In the following, we will discuss only the effects of ISL resonant impactors and BSL planetesimals assuming a simple cratering regime (i.e. no fracture creation and no weakening of the internal structure of the two asteroids).\\
As is straightforward to see, BSL impactors contribute significantly to the crustal removal only in the scenario where Jupiter formed at its present position, where their contribution is of the same order of magnitude as that of ISL resonant impactors (see Tables \ref{table3} and \ref{table4}). For Ceres, the total excavated depth in this scenario is $\approx 20$ km, while for Vesta the total excavated depth is $\approx15$ km. For increasing values of the Jovian displacement, on Ceres we noted an almost linear increase in the excavated depth, which reaches a value of $\approx110$ km in the scenario where Jupiter migrated by $1$ AU (see Table \ref{table4}). This depth implies an excavated volume equal to $90\%$ the present volume of Ceres, which is equivalent to assuming that the original mass of Ceres was about twice the present one. Interestingly, in the same scenario we noticed that the Jovian bombardment collisionally ablates Vesta, removing an amount of material actually twice as great as the mass of the asteroid (see Table \ref{table3}). Moreover, in the scenario where Jupiter migrates by $0.5$ AU both asteroids are stripped of an amount of material equivalent to roughly $50\%$ their present volumes.\\
If Jupiter migrated by more than $0.5$ AU, therefore, our results indicate that to survive the early bombardment caused by the giant planet the primordial Vesta and Ceres should have been between $150-300\%$ their present sizes, even under the simplifying assumption of a simple excavation regime.
% This is a consequence of the ISL resonant impactors produced by the $2:1$ resonance, which are also responsible for the sharp increase in the excavated depth on Vesta in the migration scenario where Jupiter formed $0.5$ AU farther away from the Sun (see Table \ref{table3}).

\subsection{Effects of a slower migration of Jupiter}\label{slowmigration}

\begin{figure*}
 \centering
 \includegraphics[angle=-90,width=17.5cm]{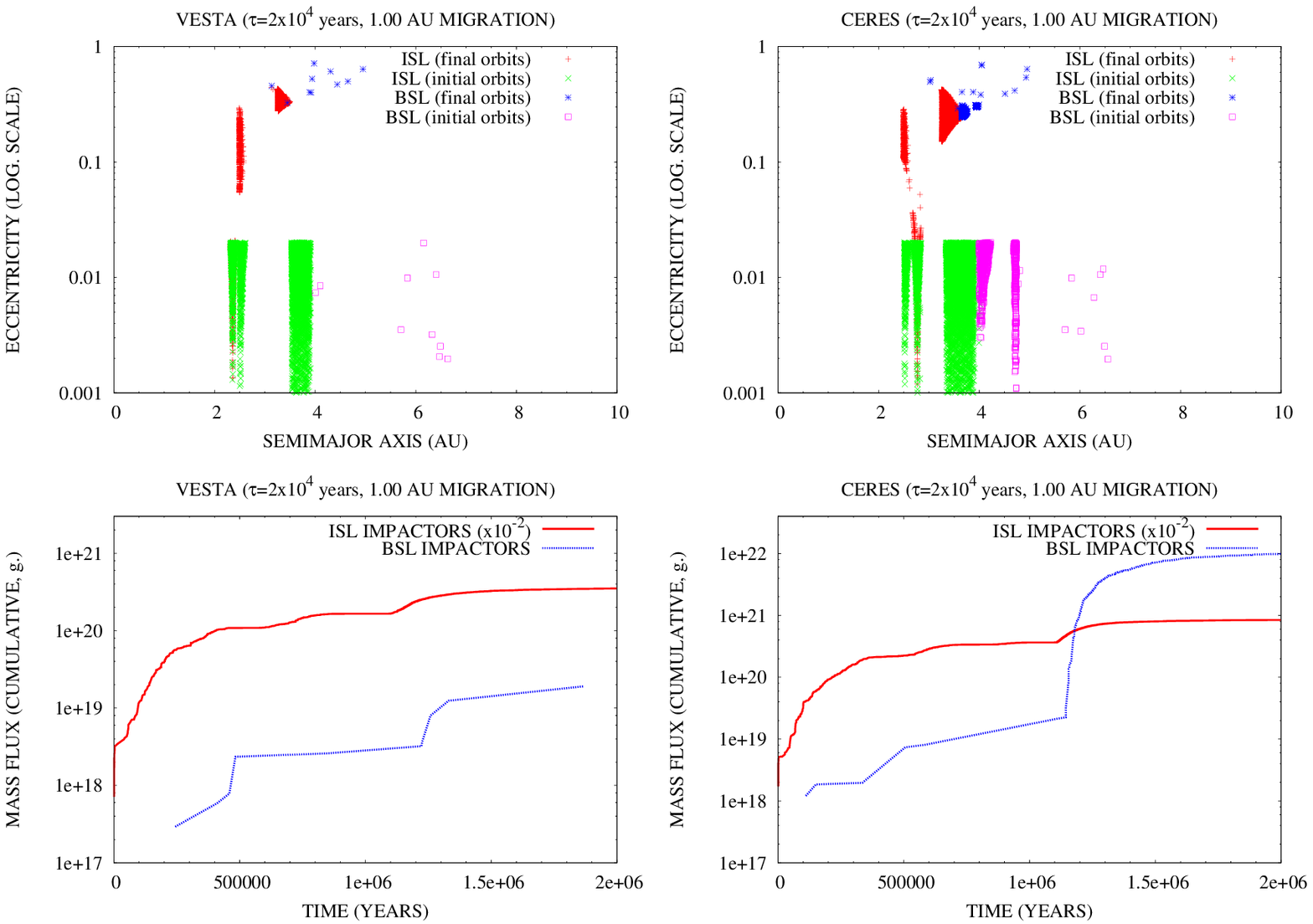}
 \caption{Jovian early bombardment for Jupiter migrating with an e-folding time of $2.5\times10^{4}$ years. Upper plots: semilogarithmic plots of the orbital elements of the impactors on Vesta (left) and Ceres (right) in the $a-e$ plane. Red and green symbols represent respectively the final (at impact) and initial orbits of ISL impactors, while the blue and magenta symbols represent respectively the final and initial orbits of BSL impactors. As in Figs. \ref{fig1} and \ref{fig2}, these plots show only the dynamical classes of impactors recorded in the simulations: they are not normalised to the real disk population. Lower plots: cumulative distribution of the mass flux over time for Vesta (left) and Ceres (right). The red curve is related to ISL impactors, the blue one to BSL impactors. The mass flux of ISL impactors in all plots has been divided by a factor $100$ to enhance the readability and facilitate the comparison of the plots.}\label{fig17}
\end{figure*}
\begin{figure*}
 \centering
 \includegraphics[angle=-90,width=17.5cm]{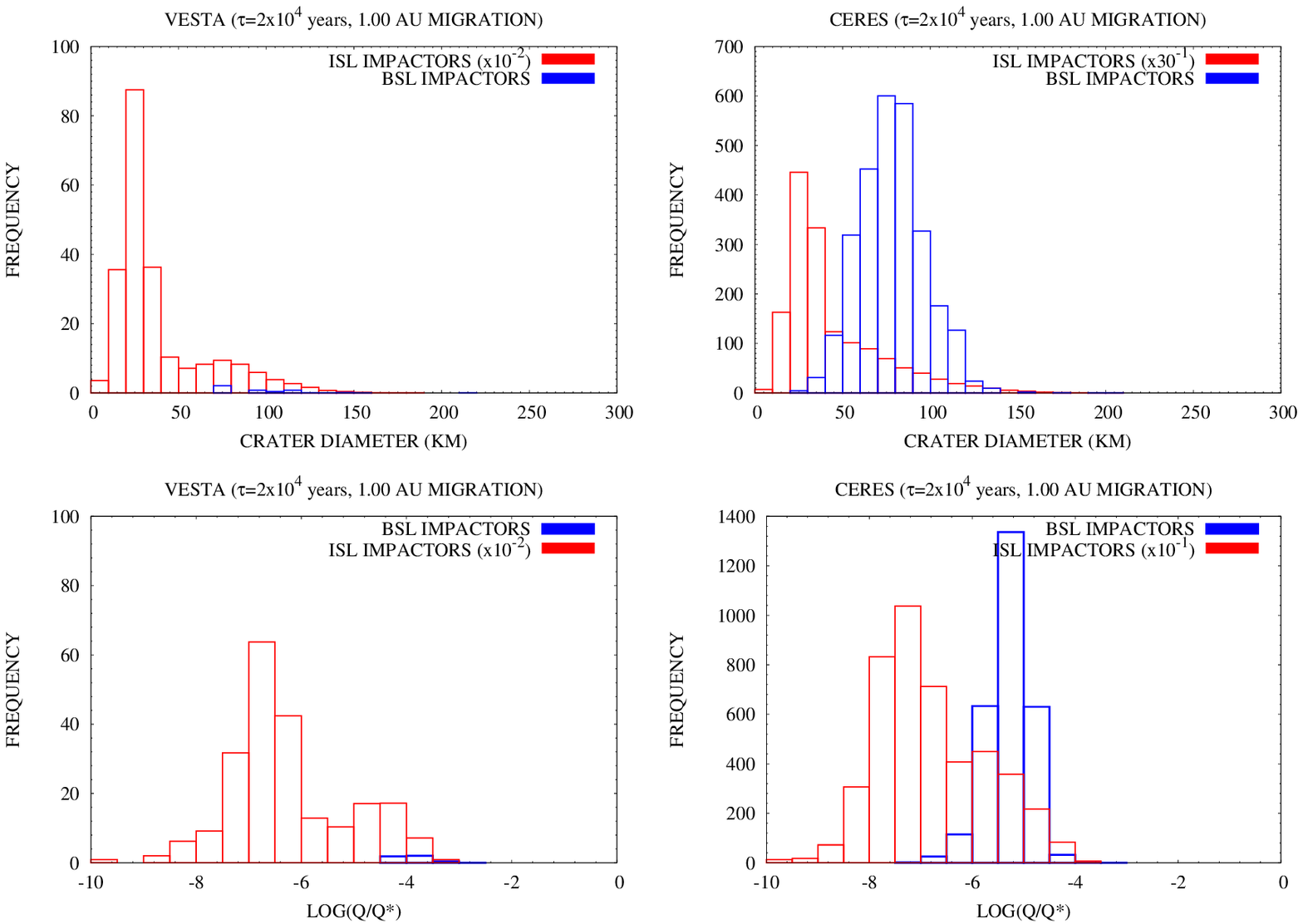}
 \caption{Jovian early bombardment for Jupiter migrating with an e-folding time of $2.5\times10^{4}$ years. Upper plots: normalised frequency versus crater size histograms of the impacts on Vesta (left) and Ceres (right). Red bars are those related to ISL impactors, blue ones to BSL impactors. The frequency of ISL impactors has been divided by a factor $100$ for Vesta and $30$ for Ceres to enhance the readability of the plots. Lower plots: normalised frequency versus impact energy (in units of the dispersal energy $Q^{*}_{D}$) histograms of the impacts on Vesta (left) and Ceres (right). Red bars are those related to ISL impactors, blue ones to BSL impactors. The frequency of ISL impactors has been divided by a factor $100$ for Vesta and $10$ for Ceres to enhance the readability of the plots.}\label{fig18}
\end{figure*}

As we discussed in Sect. \ref{jupiter}, the radial displacement of Jupiter is governed by Eq. \ref{radiallaw} and is characterised by an e-folding time $\tau_g=5\times10^3$ years, which implies a migration timescale consistent with the indications of theoretical models. However, in order to evaluate the dependence of the results on our assumption on the migration timescale, we considered an additional case respect to our four reference scenarios. We forced Jupiter to migrate inward again by $1$ AU but with an e-folding time five times as large as in our original simulations, i.e. $\tau_g=2.5\times10^4$ years. This is equivalent to assuming a value of $\tau_{M}= a/\dot{a} \approx 4\times10^5$ years at $5.2$ AU, which is consistent with the upper boundary to the migration timescale reported by \cite{dkh03}. A summary of the results for both Vesta and Ceres is given by Figs. \ref{fig17} and \ref{fig18}.\\
As can be easily seen by comparing the left-hand side plots in Figs. \ref{fig17} and \ref{fig18} with the bottom-right ones of Figs. \ref{fig1}, \ref{fig9}, \ref{fig13} and \ref{fig15}, the change in the migration timescale does not affect in any significant way the collisional evolution of Vesta respect to our reference case with Jupiter migrating by $1$ AU. The same hold true for the ISL impactors on Ceres, as can be seen by comparing the right-hand side plots in Figs. \ref{fig17} and \ref{fig18} with the bottom-right ones of Figs. \ref{fig2}, \ref{fig10}, \ref{fig14} and \ref{fig16}. The flux of BSL impactors on Ceres, on the contrary, is enhanced by a factor $4$ respect to the reference case. This enhancement, however, affects the results we previously discussed only from a quantitative point of view, not qualitatively. The major role in the collisional evolution of Ceres is still played by ISL impactors, whose flux during the bombardment is an order of magnitude higher than that of their BSL counterparts.\\
While the link between the Jovian migration rate and efficiency of the giant planet in exciting BSL resonant impactors needs further investigation, as a consequence of this additional run we feel confident that the global picture concerning the collisional evolution and the survival of both Vesta and Ceres does not depend on our assumptions on the migration rate of Jupiter.

\subsection{Collisional evolution in a turbulent disk}\label{turbulence}

\begin{figure*}
 \centering
 \includegraphics[width=17.5cm]{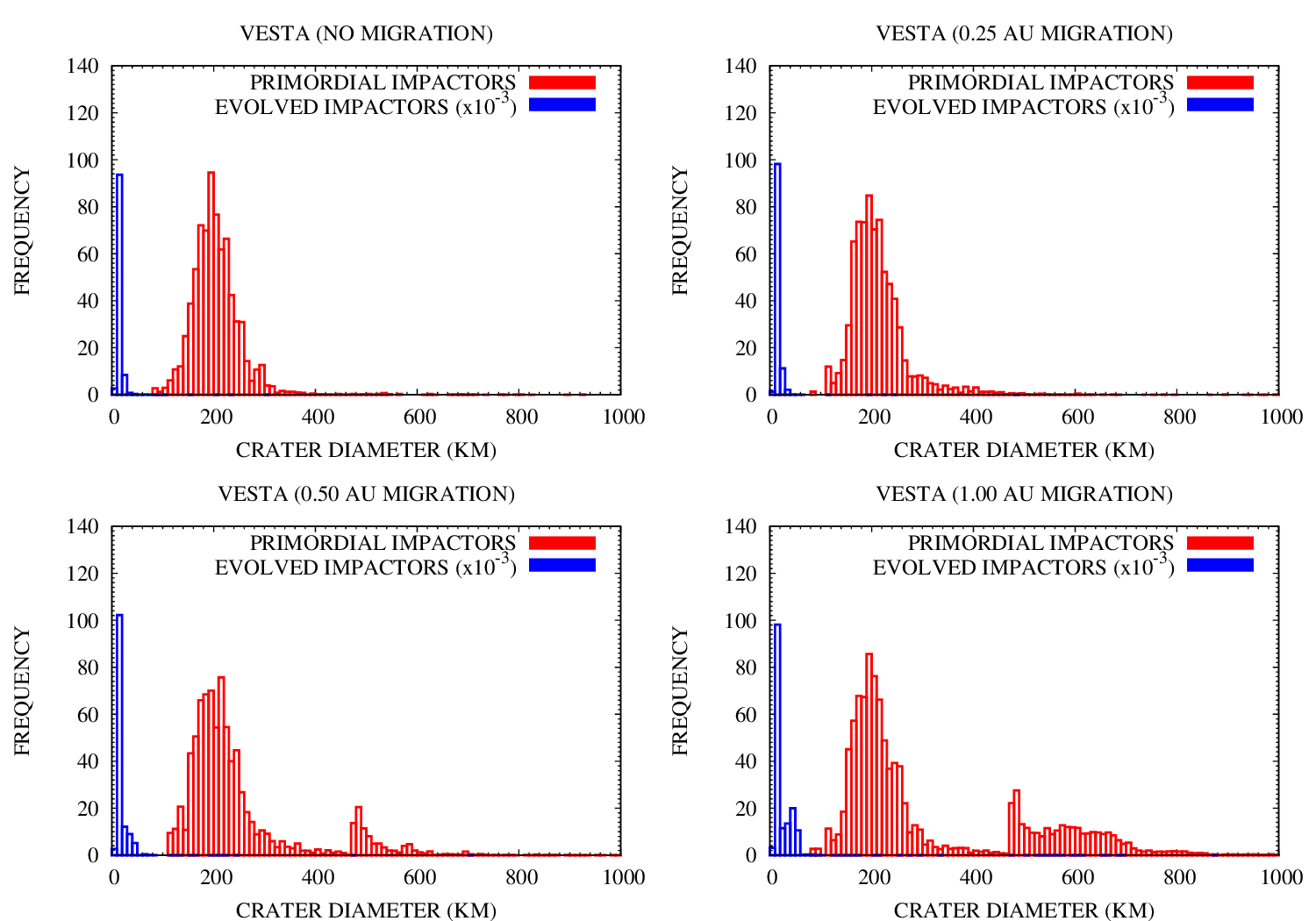}
 \caption{Normalised frequency versus crater size histograms of the impacts on Vesta computed using the primordial (red bars) and collisionally evolved (blue bars) size-frequency distributions in the inner Solar System from Fig. $8a$ in \citet{mea09}. The frequency of collisionally evolved impactors in all plots has been divided by a factor $1000$ to enhance the readability of the plots. Note that the craters whose size is comparable to or greater than the diameter of the asteroid are reported only to show the number of highly energetic impacts, since Eq. \ref{craterlaw} is not appropriate for such cases. As can be easily seen from the plots, the collisionally history of Vesta due to the primordial size-frequency distribution would result in the destruction of the asteroid. The collisionally evolved size-frequency distribution would result in a impact history more similar to the one obtained for the quiescent disk (see Fig. \ref{fig13}), but with a higher cumulative probability of impacts capable to destroy to asteroid. The survival and global impact erosion outcomes of the models are summarised in Table \ref{table5}.}\label{fig19}
\end{figure*}
\begin{figure*}
 \centering
 \includegraphics[width=17.5cm]{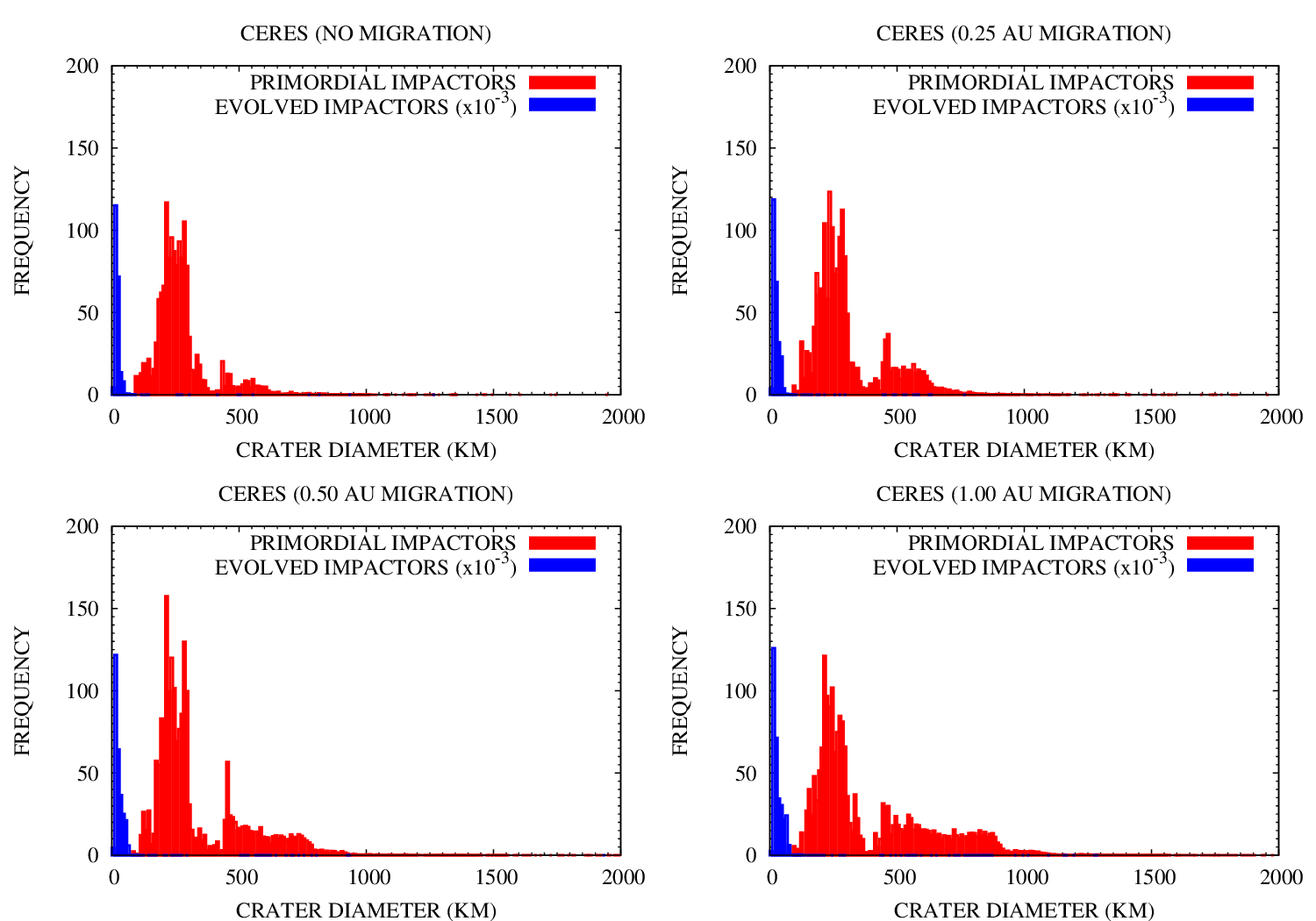}
 \caption{Normalised frequency versus crater size histograms of the impacts on Ceres computed using the primordial (red bars) and collisionally evolved (blue bars) size-frequency distributions in the inner Solar System from Fig. $8a$ in \citet{mea09}. The frequency of collisionally evolved impactors in all plots has been divided by a factor $1000$ to enhance the readability of the plots. Note that the craters whose size is comparable to or greater than the diameter of the asteroid are reported only to show the number of highly energetic impacts, since Eq. \ref{craterlaw} is not appropriate for such cases. As in the case of Vesta in Fig. \ref{fig19}, the collisionally history of Ceres due to the primordial size-frequency distribution would result in the destruction of the asteroid. Again, the collisionally evolved size-frequency distribution would result in a impact history similar to the one obtained for the quiescent disk (see Fig. \ref{fig14}), but with a higher cumulative probability of impacts capable to destroy to asteroid. The survival and global impact erosion outcomes of the models are summarised in Table \ref{table6}.}\label{fig20}
\end{figure*}
\begin{figure*}
 \centering
 \includegraphics[width=17.5cm]{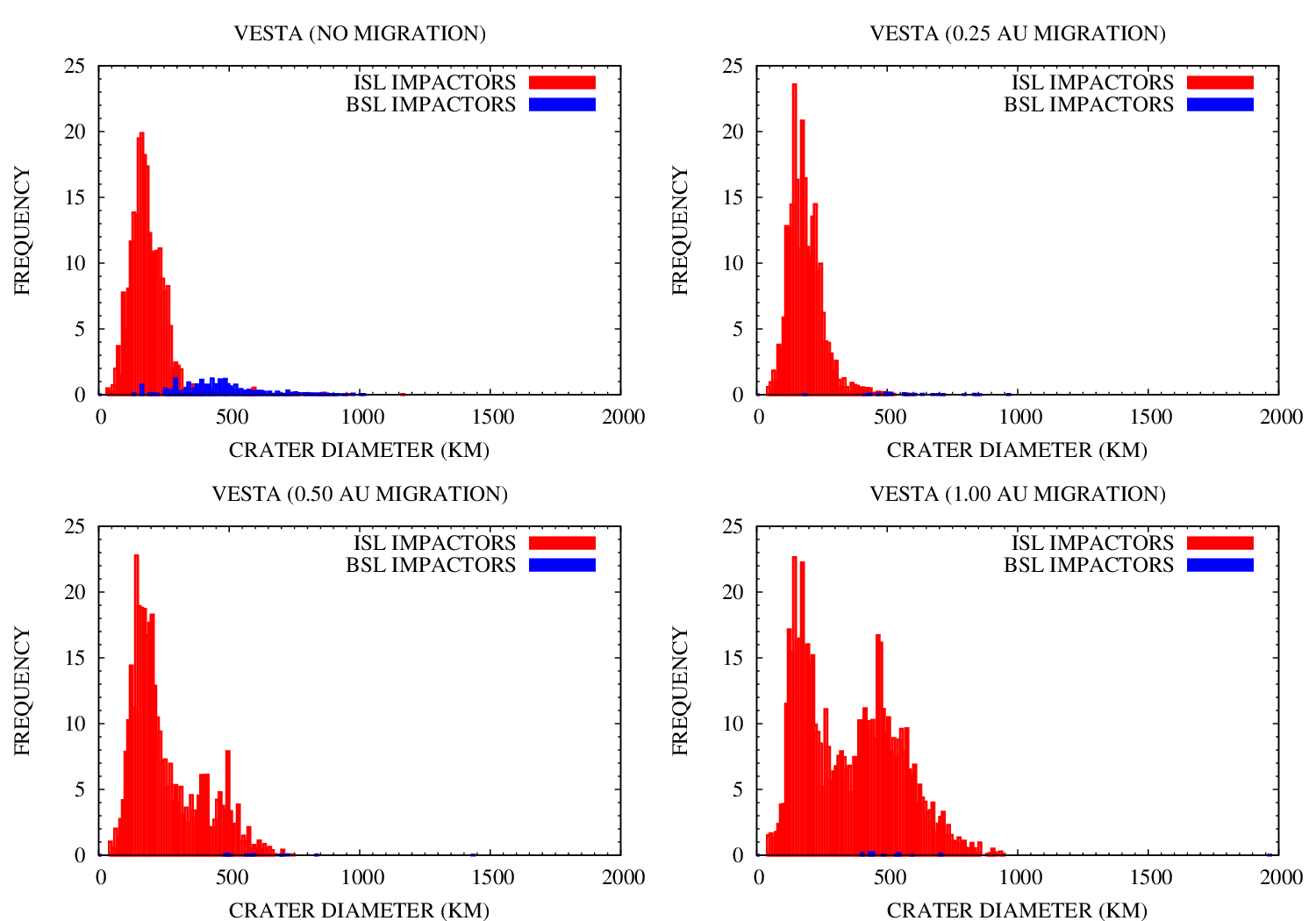}
 \caption{Normalised frequency versus crater size histograms of the impacts on Vesta computed using the diameter-heliocentric distance relationship for planetesimals from Fig. $14$ in \citet{cha10}: red bars are those related to bodies formed in the region of ISL impactors, blue ones to bodies formed in the region of BSL impactors. Note that the craters whose size is comparable to or greater than the diameter of the asteroid are reported only to show the number of highly energetic impacts, since Eq. \ref{craterlaw} is not appropriate for such cases. As can be easily seen from the plots and is summarised in Table \ref{table5}, the collisional history of Vesta in such a disk of primordial planetesimals would result in the destruction of the asteroid.}\label{fig21}
\end{figure*}
\begin{figure*}
 \centering
 \includegraphics[width=17.5cm]{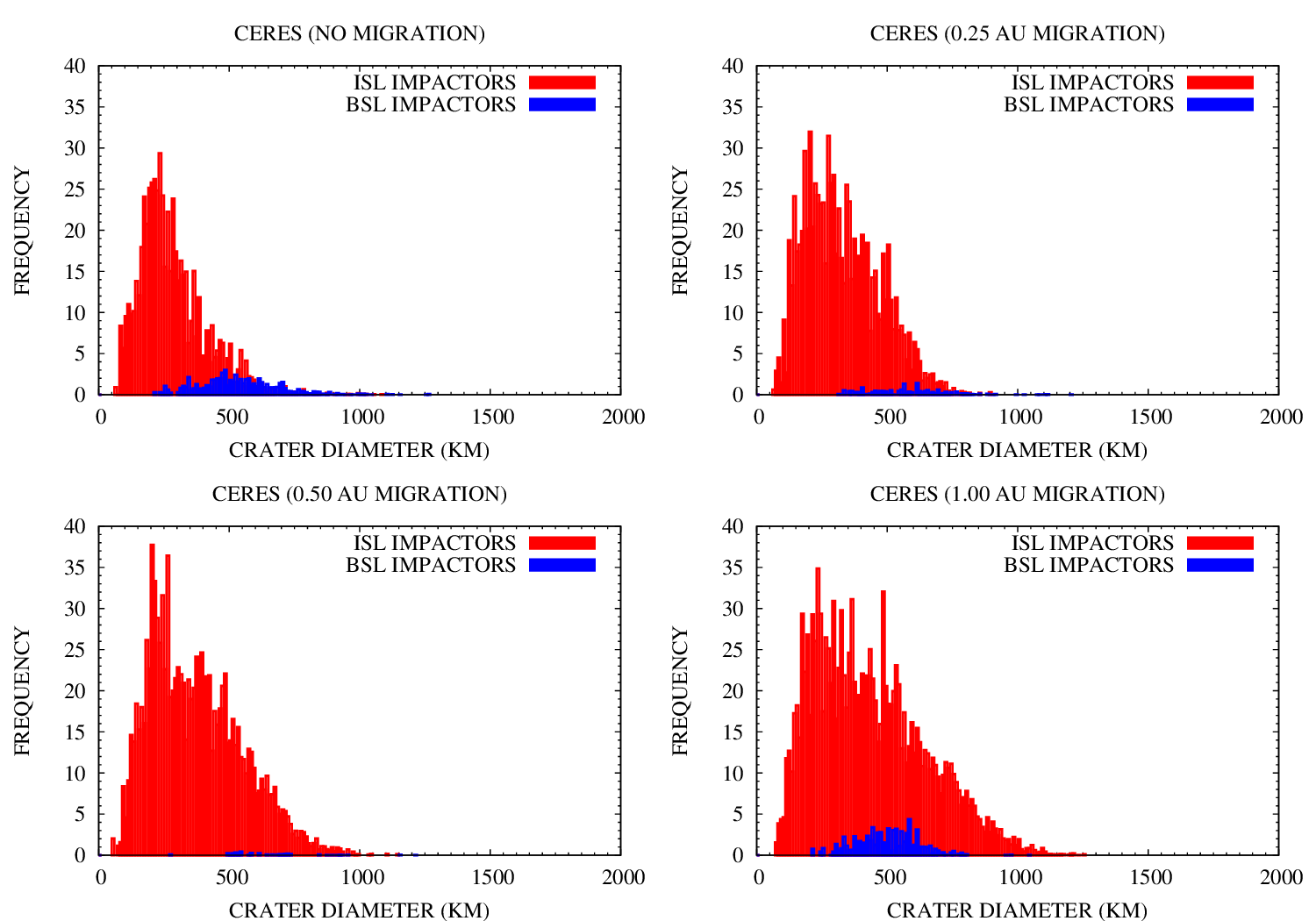}
 \caption{Normalised frequency versus crater size histograms of the impacts on Ceres computed using the diameter-heliocentric distance relationship for planetesimals from Fig. $14$ in \citet{cha10}: red bars are those related to bodies formed in the region of ISL impactors, blue ones to bodies formed in the region of BSL impactors. Note that the craters whose size is comparable to or greater than the diameter of the asteroid are reported only to show the number of highly energetic impacts, since Eq. \ref{craterlaw} is not appropriate for such cases. As we pointed out in Fig. \ref{fig21} for Vesta and is summarised in Table \ref{table6}, the collisional history of Ceres in such a disk of primordial planetesimals would result in the destruction of the asteroid.}\label{fig22}
\end{figure*}
\begin{figure*}
 \centering
 \includegraphics[width=17.5cm]{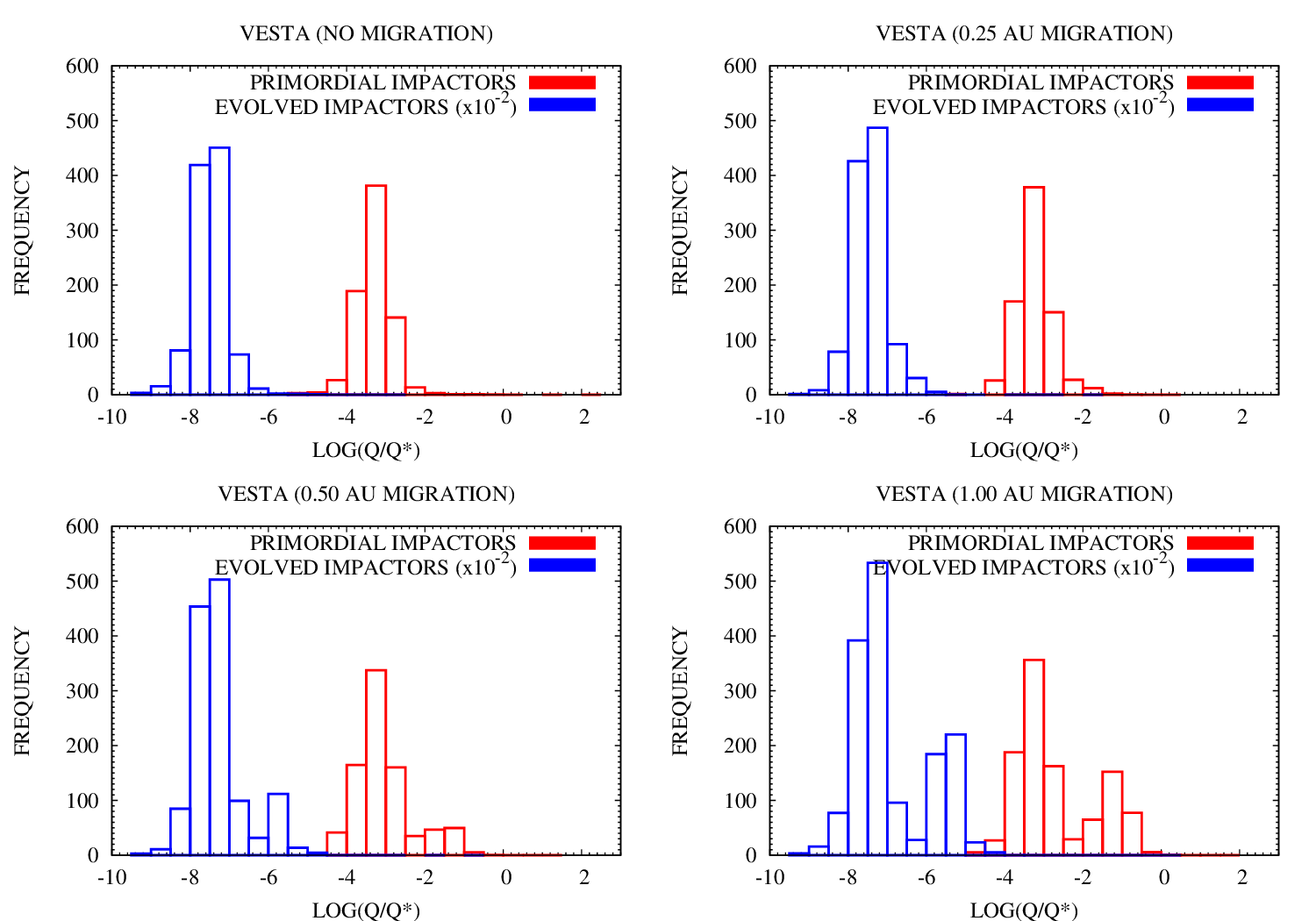}
 \caption{Normalised frequency versus impact energy (in units of the dispersal energy $Q^{*}_{D}$) histograms of the impacts on Vesta computed using the primordial (red bars) and collisionally evolved (blue bars) size-frequency distributions in the inner Solar System from Fig. $8a$ in \citet{mea09}. The frequency of the collisionally evolved impactors in all plots has been divided by a factor $100$ to enhance the readability of the plots. As can be seen in the histograms referring to the primordial SFD by \citet{mea09}, highly energetic impacts ($Q\geq10^{-2}Q^{*}_{D}$) are abundant in all scenarios where Jupiter migrates.}\label{fig23}
\end{figure*}
\begin{figure*}
 \centering
 \includegraphics[width=17.5cm]{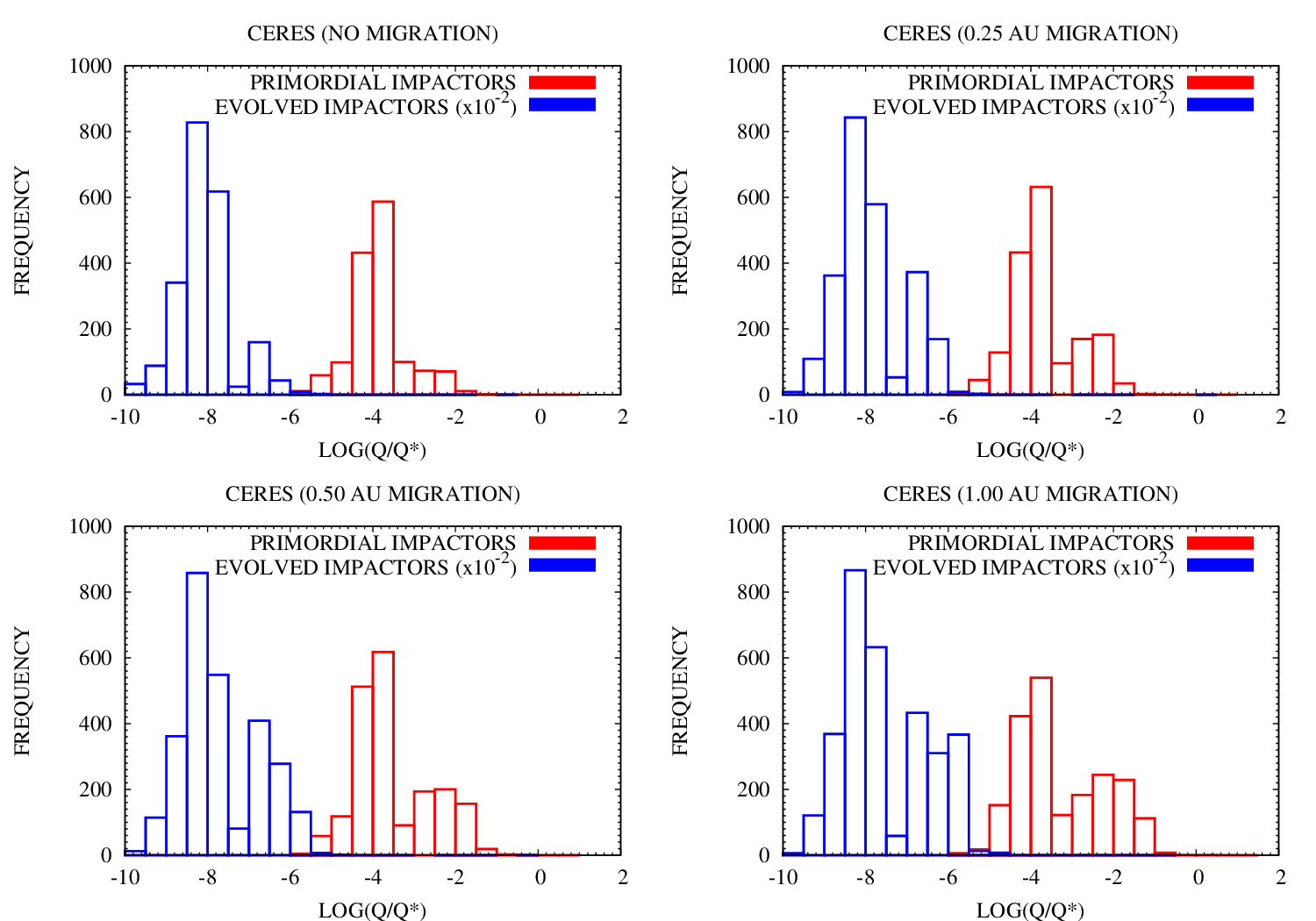}
 \caption{Normalised frequency versus impact energy (in units of the dispersal energy $Q^{*}_{D}$) histograms of the impacts on Ceres computed using the primordial (red bars) and collisionally evolved (blue bars) size-frequency distributions in the inner Solar System from Fig. $8a$ in \citet{mea09}. The frequency of collisionally evolved impactors in all plots has been divided by a factor $100$ to enhance the readability of the plots. When considering the primordial SFD by \citet{mea09}, highly energetic impacts ($Q\geq10^{-2}Q^{*}_{D}$) are abundant in all scenarios where Jupiter migrates as in the case of Vesta (see Fig. \ref{fig23}).}\label{fig24}
\end{figure*}
\begin{figure*}
 \centering
 \includegraphics[width=17.5cm]{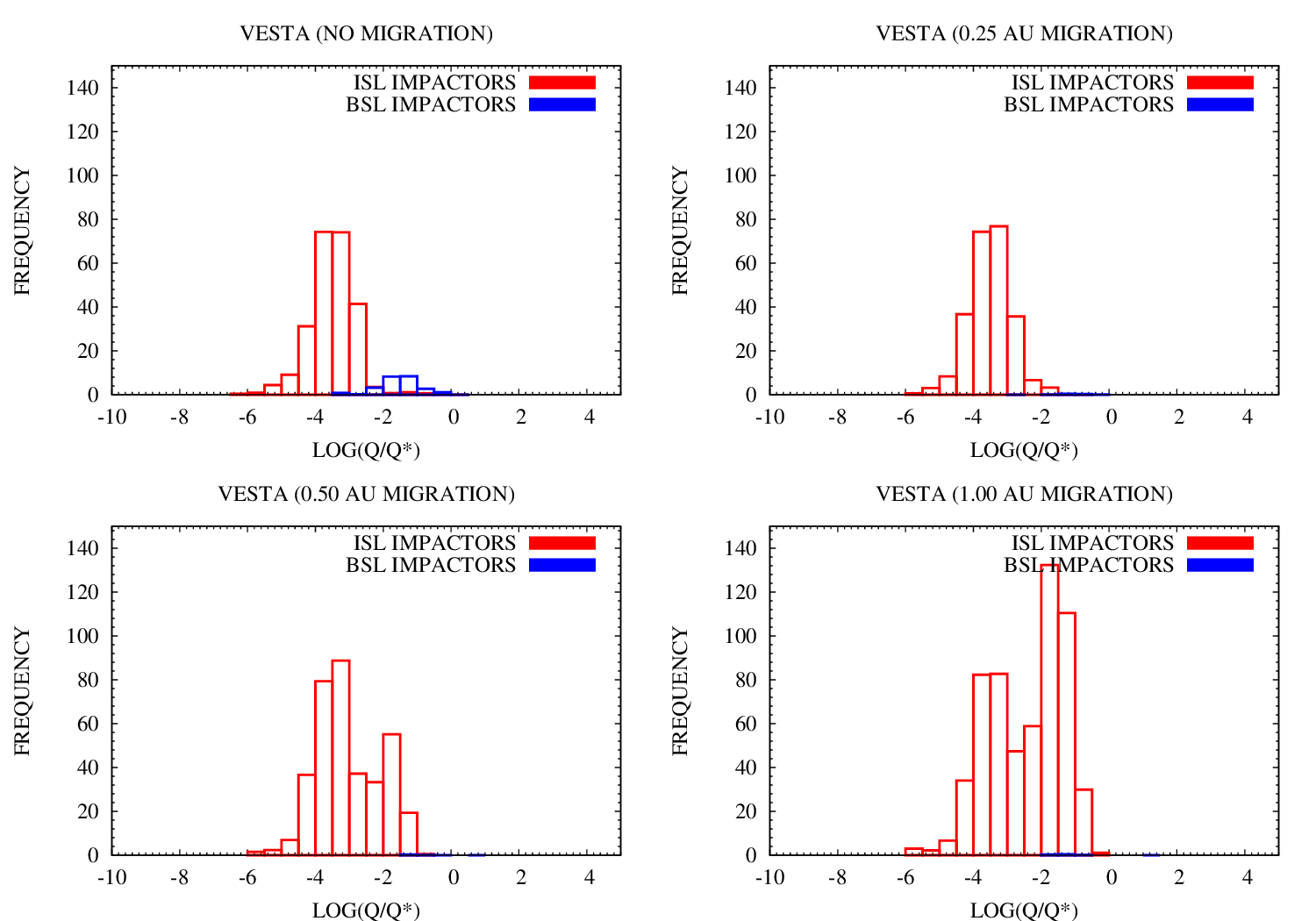}
 \caption{Normalised frequency versus impact energy (in units of the dispersal energy $Q^{*}_{D}$) histograms of the impacts on Vesta computed using the diameter-heliocentric distance relationship for planetesimals from Fig. $14$ in \citet{cha10}. Red bars are those related to bodies formed in the region of our original ISL impactors ($2-4$ AU), blue ones to bodies formed in the region of our original BSL impactors ($4-10$ AU). Highly energetic impacts ($Q\geq10^{-2}Q^{*}_{D}$) are present in all scenarios and are particularly abundant when Jupiter migrates by $0.5$ AU or more.}\label{fig25}
\end{figure*}
\begin{figure*}
 \centering
 \includegraphics[width=17.5cm]{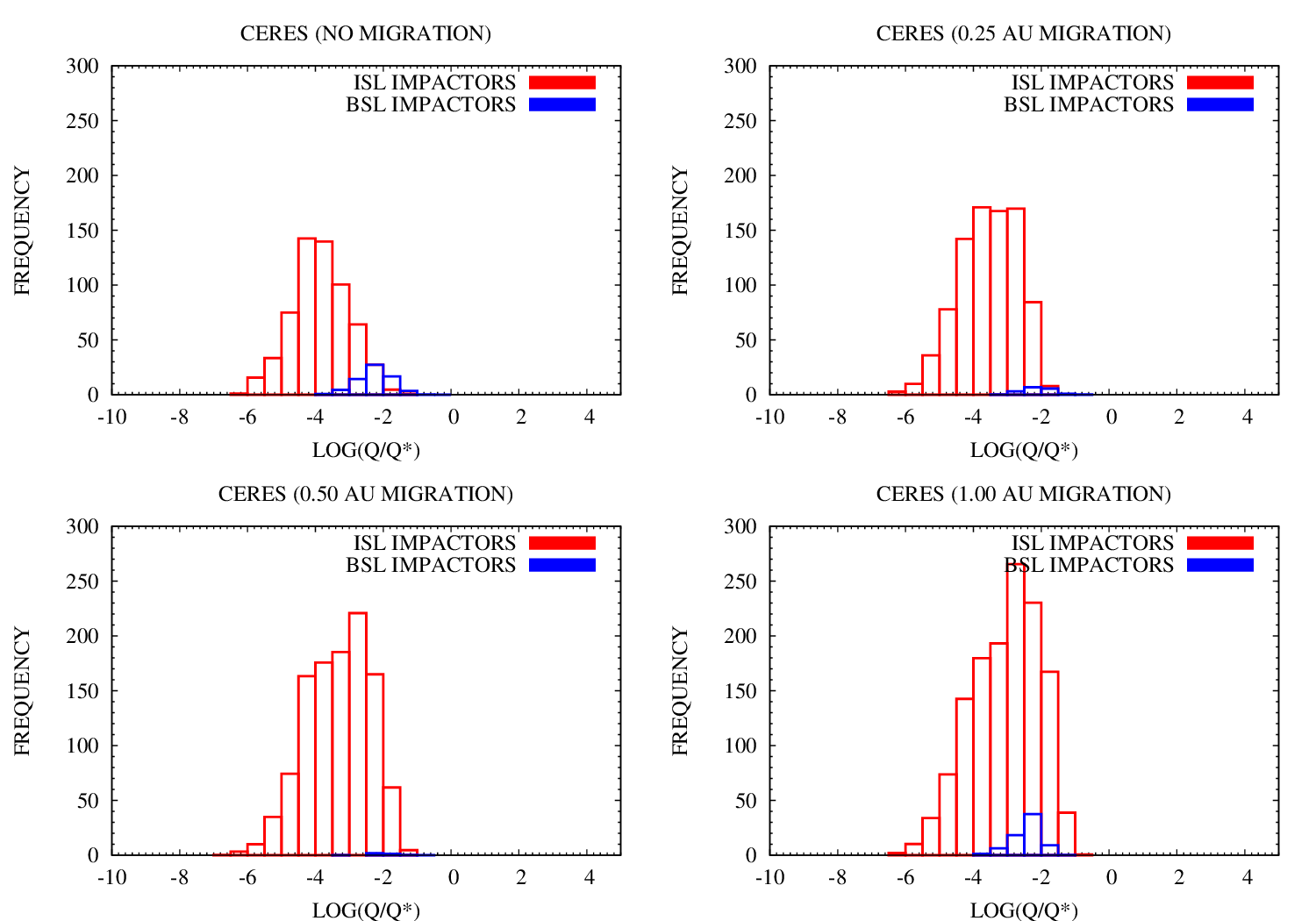}
 \caption{Normalised frequency versus impact energy (in units of the dispersal energy $Q^{*}_{D}$) histograms of the impacts on Ceres computed using the diameter-heliocentric distance relationship for planetesimals from Fig. $14$ in \citet{cha10}. Red bars are those related to bodies formed in the region of our original ISL impactors ($2-4$ AU), blue ones to bodies formed in the region of our original BSL impactors ($4-10$ AU). Similarly to the case of Vesta in Fig. \ref{fig25}, highly energetic impacts ($Q\geq10^{-2}Q^{*}_{D}$) are present in all scenarios and are particularly abundant when Jupiter migrates by $0.5$ AU or more.}\label{fig26}
\end{figure*}

As we described in Sect. \ref{disk}, the size distribution of the planetesimals in our simulations has been derived under the assumption that these bodies formed by gravitational instability of the dust in a quiescent disk \citep{saf69,gaw73}. According to \cite{cm81} and taking into account the change in density we assumed to take place across the Snow Line, the average diameters of such planetesimals roughly vary between $5-60$ km in the spatial range $1-40$ AU or, equivalently, between $7-25$ km in the interval $2-10$ AU.\\
Different planetesimal formation mechanisms, however, would produce different size distributions and, as a consequence, different planetesimal abundances than the one we considered. In particular, theoretical models of planetesimal formation in turbulent disks \citep{jea07,cuz08} predict primordial bodies whose average sizes exceed those contemplated by our model. Therefore, we proceeded to test the effects of different size distributions of the planetesimals populating the disk on our results. To do this, we took advantage of the results of \cite{mea09} and \cite{cha10}.\\
\cite{mea09} did not explore a specific model of planetesimal formation in quiescent or turbulent disks but instead tried to constrain the initial size-frequency distribution (SFD in the following) of planetesimals in the orbital region of the Main Asteroid Belt. Their results suggest that the best match with the present-day SFD of the Main Asteroid Belt is obtained for planetesimal sizes initially spanning $100-1000$ km (see Fig. $8$, ibid), a range consistent with their formation in a turbulent nebula. \cite{mea09} supplies two SFDs associated to this case: a first one describing the primordial SFD of the planetesimals, which spans $100-1000$ km (see Fig. $8a$, black dots, ibid), and a second, collisionally evolved one where accretion and break-up of the primordial planetesimals extended the size distribution between $5-5000$ km (see Fig. $8a$, black solid line, ibid). For each ISL impact event in our simulations, we then extracted the mass and the normalisation factor of the impacting planetesimal through simple Montecarlo extractions based on the cumulative probability distributions of the two SFDs supplied by \cite{mea09}. In using the results of \cite{mea09} to evaluate the collisional evolution of Vesta and Ceres we did not considered the contribution of BSL impactors, since the authors investigated the initial SFD only of planetesimals in the orbital region of the Main Asteroid Belt.\\
\cite{cha10} instead studied the planetesimal accretion efficiency following the model by \cite{cuz08}, where turbulence in the solar nebula act to concentrate dust particles in low vorticity regions. For this model, \cite{cha10} derived the accretion timescale and the size-heliocentric distance relationship of the planetesimals as a function of the mass and the gas to dust ratio of the solar nebula. We adopted the size distribution associated to a disk with gas surface density at $1$ AU $\sigma'_{0}=3500$ g cm$^{-2}$, gas to dust ratio $\xi'=0.3$ beyond the Snow Line and $\xi'=0.15$ inside the Snow Line, and a nebula density profile with exponent $n'_s=-1$ (see Fig. $14$, gray dot-dashed line, ibid). \cite{cha10} assumed the Snow Line being placed at $2.7$ AU for such nebula: due to our division of the disk into concentric rings, we assumed the Snow Line at $3.0$ AU when using the results of this author. As we mentioned, the results of \cite{cha10} supply the average diameter of planetesimals as a function of heliocentric distance: from Fig. $14$ (gray dot-dashed line) in \cite{cha10} we can derive the following analytical expression:
\begin{equation}\label{chambers_size}
 D=D_{0}\left( \frac{r}{1\,AU} \right)^{\beta'}
\end{equation}
where $D_{0}=70$ km is the average diameter of the planetesimals at $1$ AU and $\beta'=0.4935$. We can then derive a semi-empirical relationship, analogous to Eq. \ref{masslaw}, linking the average mass of planetesimals to heliocentric distance, i.e.
\begin{equation}\label{chambers_mass}
 \overline{m'}_{p}=\frac{\pi}{6}\rho'D_{0}^{3}\left( \frac{r}{1\,AU} \right)^{3\beta'}
\end{equation}
where $\rho'=3.0$ g cm$^{-2}$ in the ISL region and $\rho'=1.0$ g cm$^{-2}$ in the BSL region. By substituting the primed quantities to the original ones in Eqs. \ref{cumnum}, \ref{ntot} and \ref{massval}, we can thus obtain the mass and the normalisation factor for each planetesimal through the same approach we used for our original disk.\\
A summary of the results we obtained in using these three size distributions for the planetesimals in the disk are reported in Tables \ref{table5} and \ref{table6}, where we computed the total excavation depth of the Jupiter-driven bombardment after the giant planet started to accrete the nebular gas, and in Figs. \ref{fig19}-\ref{fig26}, where we showed the size distributions of the craters and the energies released by the impact events. As can be seen, the possible collisional histories computed for Vesta and Ceres using the primordial SFD from \cite{mea09} and the size distribution from \cite{cha10} described by Eq. \ref{chambers_size} are inconsistent with the survival of the asteroids to the Jovian early bombardment due to the following reasons.\\
First, impacts capable to disrupt Vesta (i.e. with $Q \geq Q^{*}_{D}$) are present in all our simulations, even if they are characterised by very low probabilities (see Figs. \ref{fig23} and \ref{fig25}). This holds true also for Ceres in the case of the primordial SFD by \cite{mea09} (see Fig. \ref{fig24}) but not in the case of the size distribution by \cite{cha10} (see Fig. \ref{fig26}). In addition, impacts delivering a significant fraction (i.e. $Q \geq 10^{-2}Q^{*}_{D}$) of the dispersal energies $Q^{*}_{D}$ of Vesta and Ceres are abundant (as shown in Figs. \ref{fig23}-\ref{fig26}), especially in those scenarios where Jupiter migrated by $0.5$ AU or more. The cumulative effects of these energetic impacts, not accounted for in our simplified calculations, would likely result in the weakening of the internal structure of the asteroids and in their eventual destruction.\\
Second, even if Vesta and Ceres could survive the Jovian early bombardment without being disrupted, the cumulative volume excavated by the impactors would result in the ablation of the two asteroids even assuming a simple cratering regime (see Tables \ref{table5} and \ref{table6}). Moreover, for those impacts producing craters whose size is comparable to the diameter of the relevant asteroid (see Figs. \ref{fig19}-\ref{fig22}), the cratering regime described by Eq. \ref{craterlaw} is not appropriate. According to Eq. $8$ by \cite{baa99}, impacts characterised by $Q\sim10^{-2}Q^{*}_{D}$ would remove from the target asteroid a mass of the order of $\sim1\%$ for impact velocities of about $3$ km s$^{-1}$ and of $15-20\%$ for impact velocities of about $5$ km s$^{-1}$ (not taking into account the cumulative effects of such energetic impacts). As a consequence, the survival of Vesta and Ceres in disks where the size distribution of the planetesimals is governed by the primordial SFD from \cite{mea09} and Eq. \ref{chambers_size} as in \cite{cha10} appears as extremely unlikely. Even in the scenario where the bombardment is less efficient (i.e. Jupiter migrating inward by $0.25$ AU), Ceres would be collisionally ablated while Vesta would be stripped of a mass equivalent to about $80\%$ its present one.\\
The case of the collisionally evolved SFD from \cite{mea09} is different. In such collisionally evolved SFD, about $84\%$ of the ISL population is represented by planetesimals whose diameter is $\sim5$ km and about $97\%$ of said population is characterised by diameters in the range $5-10$ km, i.e. our original size range in the ISL orbital region. As is shown in Figs. \ref{fig19} and \ref{fig20} for Vesta and Ceres respectively, the crater size distributions computed using this SFD are similar to the ones we obtained for our original cases. The bulk of the craters is characterised by diameters lower than $50-75$ km depending on the migration scenario considered, while the values of the impact energy $Q$ are generally $Q\leq10^{-4}Q^{*}_{D}$ for Vesta and $Q\leq10^{-5}Q^{*}_{D}$ for Ceres. Respect to the case of planetesimals formed in a quiescent disk, the collisional histories of Vesta and Ceres estimated using this second SFD from \cite{mea09} are characterised by higher probabilities of highly energetic impacts, with energies $Q\geq10^{-2}Q^{*}_{D}$ (see Figs. \ref{fig23} and \ref{fig24}) and producing craters with diameters bigger than $150-200$ km according to Eq. \ref{craterlaw} (see Figs. \ref{fig19} and \ref{fig20}). However, even neglecting the contribution of BSL impactors, the collisional histories computed for Vesta using such SFD suggest that the asteroid could survive the bombardment only if Jupiter migrated by less than $0.5$ AU. Even in this case, the equivalent volume excavated by impacts is about twice than that estimated in the case of our original computations (see Table \ref{table5}). According to our results, the survival of Ceres is even less plausible: being located between the $3:1$ and the $2:1$ resonances with Jupiter, the asteroid is collisionally ablated by the Jovian bombardment in three scenarios over four. Moreover, even if Jupiter did not migrate the volume excavated would be of the order of half the present one of the asteroid (see Table \ref{table6}), thus implying a more massive primordial Ceres.

\begin{table*}
\caption{Collisional erosion of Vesta due to planetesimals formed in a turbulent disk in the four migration scenarios of Jupiter considering only the post-core bombardment. N.B.: the excavated depth is estimated is all cases where Vesta is not collisionally ablated assuming that the final radius of the asteroid is the present one.}
\centering
\begin{tabular}{ccccc}
\hline
 \textbf{Migration scenario} & & \textbf{Excavation depth} & \textbf{(turbulent disk)} & \\
\hline
 & Primordial SFD &  Evolved SFD & ISL impactors & BSL impactors\\
 & \citep{mea09} & \citep{mea09} & \citep{cha10} & \citep{cha10}\\
\hline
No migration & Ablation & $9.847$ km & $55.88$ km & Ablation \\
$0.25$ AU & Ablation & $14.12$ km & $46.13$ km & $16.69$ km\\
$0.50$ AU & Ablation & Ablation & Ablation & $11.62$ km\\
$1.00$ AU & Ablation & Ablation & Ablation & $15.94$ km\\
\hline
\end{tabular}
\label{table5}
\end{table*}
\begin{table*}
\caption{Collisional erosion of Ceres due to planetesimals formed in a turbulent disk in the four migration scenarios of Jupiter considering only the post-core bombardment. N.B.: the excavated depth is estimated is all cases where Ceres is not collisionally ablated assuming that the final radius of the asteroid is the present one.}
\centering
\begin{tabular}{ccccc}
\hline
 \textbf{Migration scenario} & & \textbf{Excavation depth} & \textbf{(turbulent disk)} & \\
\hline
 & Primordial SFD &  Evolved SFD & ISL impactors & BSL impactors\\
 & \citep{mea09} & \citep{mea09} & \citep{cha10} & \citep{cha10}\\
\hline
No migration & Ablation & $69.54$ km & Ablation & Ablation \\
$0.25$ AU & Ablation & Ablation & Ablation & $68.54$ km\\
$0.50$ AU & Ablation & Ablation & Ablation & $21.28$ km\\
$1.00$ AU & Ablation & Ablation & Ablation & Ablation\\
\hline
\end{tabular}
\label{table6}
\end{table*}
% 
% \subsection{}
% 
\section{Implications for Vesta, Ceres and the early Solar System}\label{implications}

The picture that our results supply of the time of Jupiter's formation clearly states that the formation of the giant planet has been one of the milestones in the evolution of the early Solar System and in particular of the region of the present Main Asteroid Belt. In the following, we will discuss the consequences of our results for Jupiter, Vesta, Ceres and the early Solar System.

\subsection{Planetesimals in the primordial Solar System}\label{planetesimals_disc}

The results we described in Sect. \ref{impact_features} and in Sect. \ref{turbulence}, in particular the collisional erosion and the likelyhood of survival of Vesta and Ceres to the Jupiter-induced bombardment depicted by Tables \ref{table3}-\ref{table6}, give us interesting constrains on the population of planetesimals in the Solar Nebula.\\
As we explained in Sect. \ref{disk}, in our simulations we initially assumed that all the mass of the disk was accounted for by planetesimals formed by gravitational instability of the dust in a quiescent nebula and whose diameters spanned the range $7-25$ km. According to our results, the Jovian early bombardment would cause an intense cratering and a significant erosion of the surfaces of Vesta and Ceres but the two asteroids would survive it if the giant planet migrated by less than $0.5$ AU (see discussion in Sect. \ref{jupiter_disc}).\\
We then tested the effects of the Jovian early bombardment in disks characterised by different size distributions of the planetesimals: specifically, we took advantage of the results of \cite{mea09} and \cite{cha10}. According to our results (see Tables \ref{table5} and \ref{table6}), the survival of Vesta and Ceres to the Jovian early bombardment is extremely unlikely in disks populated by massive planetesimals like those expected to form in turbulent disks (e.g. the size distribution by \cite{cha10} and the primordial SFD by \cite{mea09}).\\
Disks where the population of planetesimals is dominated by small bodies (i.e. $D\leq20$ km) but where a significant fraction of the mass is in the form of larger objects (i.e. $D\geq1000$ km), as the one suggested by the collisionally evolved SFD by \cite{mea09}, represent a less hostile environment from the perspective of the survival of Vesta and Ceres to the Jovian early bombardment. However, the abundances of planetesimals predicted by \cite{mea09} cause a bombardment on Ceres four times as erosive than the one estimated for our original disk, thus making its survival extremely unlikely. Moreover, the collisional erosion of Vesta would be about twice as large as the one we estimated for our original disk, therefore making the survival of Vesta even less plausible if the giant planet migrated by $0.5$ AU or more (see discussion in Sect. \ref{jupiter_disc}).\\
Before proceeding, we must point out that, strictly speaking, our results refer to a disk of planetesimals which did not undergo any process of mass depletion like the one implied by the ``native embryos'' model \citep{wet92,pmc01,omb07}. Again, we stress that the mass depletion suggested by the Nice Model \citep{gom05} is located several $10^{8}$ years in the future respect to the timeframe we explored and thus has no implications for the present work. According to \cite{pmc01}, the presence of planetary embryos in the region of the Main Asteroid Belt does not cause a significant depletion of the planetesimals if Jupiter and Saturn are not present. This implies that the mass depletion process should be active during the last $10^{6}$ years of our simulations, likely with a lower efficiency than reported by \cite{omb07} since we assumed that the formation of Saturn is delayed by several $10^{5}$ years respect to that of Jupiter. According to the results of \cite{omb07} for the case where Jupiter and Saturn were initially on circular orbits (CJS, ibid), the combined perturbations of the giant planets and the planetary embryos should account for a depletion of about $10\%$ of the original planetesimals on this timespan. As a consequence of the short timescale the Jovian early bombardment acts on (few $10^{5}$ years), the depletion mechanism of the ``native embryos'' model should not affect our results significantly. It must be noted, moreover, that a higher efficiency of the depletion mechanism on a timespan of $10^{6}$ years, like the one estimated by \cite{omb07} for the case where Jupiter and Saturn were on orbits similar to their present ones (i.e. $\sim30\%$), would favour the survival of Vesta and Ceres without qualitatively changing the global picture supplied by our results. In the case of disks with a high abundance of small planetesimals and a significant fraction of the mass in the form of large planetesimals, e.g. the one described by the collisionally evolved SFD from  \cite{mea09}, depletion efficiencies as high as the one estimated by \cite{pmc01}, i.e. $\sim70\%$, could be required to ensure the survival of Ceres.
% Even assuming a depletion of $\approx70\%$ like the one suggested by \cite{pmc01}, moreover, the global picture supplied by our results would hold: 

\subsection{Jupiter}\label{jupiter_disc}

According to our simulations, the extent of Jupiter's radial migration due to disk-planet interactions has major implications for the survival of Vesta and Ceres to the Jovian early bombardment. The timescale of the migration also plays a role in influencing the intensity of the bombardment (see Sect. \ref{slowmigration}), but mainly for what concerns the flux of BSL impactors. Thus, it influences the production of major craters (i.e. $D>100$ km) but not the likelyhood of the survival of Vesta and Ceres to the Jupiter-induced bombardment.\\
In the case of our original disk of planetesimals, our results set a upper limit of $0.5$ AU to Jupiter's migration. For radial displacements higher than $0.5$ AU during its formation, in fact, the sweeping of the mean motion resonances with Jupiter through the disk of planetesimals causes an enhancement in the flux of impactors so high that Vesta is effectively collisionally ablated and Ceres is stripped of a volume equivalent to $90\%$ of its present one (see Table \ref{table3}). Moreover, in the case of a radial migration of Jupiter of about $0.5$ AU, the volume excavated by the impacts on both asteroids is of the order of half their present volumes, implying the complete removal of their crustal layers. As a consequence, the radial migration of Jupiter due to disk-planet interactions should have been inferior to $0.5$ AU.\\
For a disk numerically dominated by small bodies but with a significant fraction of the mass in the form of larger objects like in the case of the collisionally evolved SFD by \cite{mea09}, the constrains on the radial migration of Jupiter implied by the survival of Vesta and Ceres are even stricter, i.e. the displacement should have been lower than $0.25$ AU. According to our results, in fact, Vesta would survive the Jovian early bombardment for displacements of the giant planet inferior to $0.5$ AU. However, being located between the $3:1$ and the $2:1$ resonances, Ceres would be subjected to a bombardment so intense that it would be collisionally ablated even for displacements of the giant planet of the order of $0.25$ AU.

\subsection{Ceres}\label{ceres_disc}

As we discussed previously in Sect. \ref{planetesimals_disc} and \ref{jupiter_disc} and is shown in Tables \ref{table4} and \ref{table6}, the survival of Ceres to the Jovian early bombardment is linked to the extent of Jupiter's migration and to the characteristics of the planetesimals populating the Solar Nebula. According to our results, the survival of Ceres to the flux of impactors coming from the $3:1$ and the $2:1$ resonances would require a limited displacement of the forming Jupiter, inferior to $0.5$ AU and possibly lower than $0.25$ AU.\\
If we consider only those scenarios where Ceres survived the bombardment, the flux of impactors would have removed a significant fraction of the original crust of the asteroid, ejecting the fragments in the Main Asteroid Belt. Even in the less collisionally active scenarios, in fact, the primordial Ceres could have been $\approx15-20\%$ bigger than it is now. Independently by the migration scenario considered, moreover, the Jovian early bombardment would have left the surface of the asteroid saturated with craters spanning up to $150$ km, with a small number of bigger craters in the range $150-300$ km.\\
It is reasonable to assume that bodies as big as Ceres populated the Main Asteroid Belt at the time of Jupiter's formation, yet we do not know if Ceres belonged to such population. The only indication in this sense is probably its mass: it is not easy, in fact, to form planetesimals as large as Ceres in the Main Asteroid Belt once Jupiter started to perturb it. If the accretion of Ceres ended before the formation of Jupiter, the surface of the asteroid could have kept records of the bombardment induced by the giant planet. If the accretion of Ceres ended at a later time, it is likely that its late phases reshaped the surface and erased every vestigial feature of this ancient bombardment.\\
Similar considerations apply also to the reshaping of the asteroid due to its thermal evolution. \cite{tea05} suggest that Ceres could be a differentiated body composed of water ice and silicate material yet, if true, we do not know if its differentiation ended before the Jovian early bombardment or if it ended later, causing a resurfacing that altered its primordial collisional features.\\
Finally, we must take into account that the surface of Ceres should have been altered by its subsequent collisional evolution through the $4$ Ga of the life of the Solar System and by the event known as the Late Heavy Bombardment. Before we can apply our results to probe the early history of Ceres, we therefore must assess to which extent later collisional or thermal events modified or erased the primordial features of its surface.

\subsection{Vesta}\label{vesta_disc}

Respect to the case of Ceres, the survival of Vesta puts less stringent constrains to the extent of Jupiter's migration: according to our simulations, the most favourable scenarios are those where the giant planet migrated by less than $0.5$ AU (see Tables \ref{table3} and \ref{table5}).\\
In those scenarios where the asteroid survived to the Jovian early bombardment, our results indicate that Vesta underwent to an intense collisional evolution and that its surface was saturated by craters spanning up to $150$ km, with a few craters of $\approx200$ km. The total amount of material excavated from Vesta is $10-20\%$ of its present volume even in the less collisionally active scenarios we considered.\\
As for Ceres, we cannot rely only on these primordial impact features to date the surface of Vesta, since they were altered or possibly erased by the subsequent $4$ Ga of collisional evolution of the Main Asteroid Belt and the passage of Vesta through the Late Heavy Bombardment. However, the case of Vesta is different from that of Ceres: thanks to its connection to HED meteorites, we know that Vesta differentiated and formed its basaltic crust in less than $4$ Ma since the formation of CAIs \citep{kei02,sco07}. Such early epoch and short timescale indicate that short-lived radionuclides like $^{26}$Al and $^{60}$Fe were the drivers of such differentiation. Preliminary modeling of the thermal evolution of Vesta indicates that its mantle would have been in a molten state for several Ma (Federico, Coradini \& Pauselli, in preparation, but see also \citet{kei02} and references therein), i.e. across the timespan of the Jovian early bombardment. Moreover, thermal models (ibid) indicate that the thickness of the crust of Vesta would have varied between a few km to about $10-20$ km in the temporal frame of interest, also depending on the amount of $^{26}$Al originally available.\\
Our results (see Tables \ref{table3} and \ref{table5}) indicate that the Jovian early bombardment should have excavated partially or completely the primordial crust of Vesta, thus creating fractures or generating uncompensated negative gravity anomalies. These would have caused effusive phenomena from the underlying mantle, in analogy with lunar maria, or the solidification of the exposed layer of the mantle and the formation of a new basaltic crust. The crystallisation epoch of these regions on the surface of Vesta would then be directly connected to the time of formation of Jupiter. Dating the crust of Vesta or the possible Vestian maria once Dawn mission will visit the asteroid would then supply the opportunity to date the Jovian early bombardment and to constrain with unprecedented accuracy the formation of Jupiter.

\section{Conclusion}\label{conclusion}

In this work we explored the collisional evolution of Vesta and Ceres at the time of Jupiter's formation, one of the milestones in the history of the early Solar System. The gravitational perturbations of the giant planet, in fact, excite the orbital resonances both inside and outside the Snow Line and trigger an early, intense bombardment in the orbital region of the Main Asteroid Belt. If Vesta or Ceres formed before Jupiter's core reached its critical mass, they would have been subject to such Jovian early bombardment.\\
In this first investigation we did not include the presence of Saturn, which is equivalent to assuming that its formation was delayed by several $10^{5}$ years respect to that of Jupiter, and we ignored the perturbing effects of gas drag and of possible planetary embryos embedded in the Solar Nebula on the dynamical evolution of the planetesimals.\\
The survival of Vesta and Ceres to this primordial bombardment depends on the characteristics of the planetesimals populating the Solar Nebula and specifically on their size distribution. Our results clearly indicate that the abundance of large planetesimals in the disk (and particularly in the region of the Main Asteroid Belt) is a critical factor for the survival of the two asteroids. If the disk of planetesimals was dominated by large bodies (i.e. $D\geq100$ km), like in the case of planetesimals formed in turbulent circumstellar disks \citep{mea09,cha10}, the two asteroids would not have survived to the Jupiter-induced bombardment. Conversely, disks dominated by small planetesimals (i.e. $D\leq20$ km), like those formed in quiescent circumstellar disks \citep{cm81} or produced by collisional evolution of larger bodies \citep{mea09}, represent more favourable environments for what it concerns the survival of Vesta and Ceres.\\
The early cratering histories of the two asteroids are extremely sensitive also to the radial migration of Jupiter due to disk-planet interactions during the phase of accretion of its gaseous envelope. According to our results, displacements of Jupiter of the order of $0.5$ AU or more would cause either an unlikely high collisional erosion of the primordial Vesta and Ceres or their destruction. Depending on the abundance of large planetesimals ($D\geq100$ km) in the disk, moreover, the survival of Ceres could imply that Jupiter migrated inward by less than $0.25$ AU. These constrains to the extent of the radial displacement of Jupiter are in agreement with the findings of other authors (see e.g. \cite{sco06} and references therein).\\
Finally, in all scenarios where they survived to the Jovian bombardment, Vesta and Ceres underwent to an intense cratering that saturated their surfaces with craters as big as $150$ km, with a tail of few bigger craters ($200-300$ km). Craters as big as the south pole impact basin on Vesta ($D\approx400$ km) are characterised in our simulations by extremely low probabilities (of the order of a few per cent). Under the simplifying assumption of a uniform distribution of the craters produced in the simple cratering regime described by Eq. \ref{craterlaw}, our results indicate that the Jovian early bombardment would have excavated a depth of about $10-15$ km on Vesta and of $20-30$ km on Ceres.\\
While it is reasonable to assume that planetesimals as big as Ceres were already present in the Main Asteroid Belt at the time Jupiter formed, we have little information on the real timescale of the accretion of Ceres. Therefore, we do not know if Ceres already completed its formation at the time of the events we simulated. Moreover, we ignore the details of the thermal evolution of the asteroid. As a consequence, we cannot rule out that, if the asteroid differentiated, its thermal evolution reshaped both the internal structure and the surface morphology, obliterating the traces of such ancient times as that covered in our simulations.\\
On the contrary, through its connection with HED meteorites we know that Vesta differentiated and formed a basaltic crust in the same temporal frame as that of Jupiter's formation. Our results suggest that the Jovian early bombardment was intense enough to excavate the Vestian crust and expose, locally or possibly even globally, the underlying mantle. Effusive phenomena would produce basaltic regions similar to lunar maria, whose epoch of crystallisation would then be directly related to that of Jupiter's formation. In all non-destructive cases resulting from our simulations, the crustal erosion would link large regions on the surface of the asteroid to the time of the Jovian early bombardment.\\
% Before concluding, we would like to point out that the Jovian early bombardment described in our model is likely a lower limit to the real event. In our simulations, in fact, we did not included the effects of gas drag on the motion of the planetesimals. By causing a radial, inward migration of the planetesimals \citep{wei77}, gas drag would replenish the mean motion resonances and therefore enhance the population of resonant impactors, which in our results acts as the major player in the collisional evolution of Vesta and Ceres. This implies that if Jupiter formed in a gas-rich nebula, instead of the gas-poor one implicitly assumed by our model, the cratering histories of Vesta and Ceres would have been more intense than the one we simulated and the constrains to Jupiter's migration could become more strict.

\section*{Acknowledgements}

The authors wish to thank the reviewer, John Chambers, for his comments and suggestions that helped to improve the manuscript and the quality of the results. D.T. also wish to thank Federico Tosi for his useful comments on the manuscript.
This research has been supported by the Italian Space Agency (ASI) through the ASI-INAF contract I/015/07/0. The computational resources used in this research have been supplied by IFSI-Rome through the project ``HPP - High Performance Planetology''.

\bsp

\label{lastpage}

\end{document}